\documentclass[sigconf, authorversion, nonacm]{acmart}

\usepackage{algorithmic}
\usepackage{amsfonts}
\usepackage{amsmath}
\usepackage{enumitem}
\usepackage{graphicx}
\usepackage{hyperref}
\usepackage{makecell}
\usepackage{multirow}
\usepackage{subcaption}
\usepackage{textcomp}
\usepackage{url}
\usepackage{xcolor}
\usepackage{xurl}

\DeclareMathOperator*{\argmin}{argmin}

\AtBeginDocument{%
  \providecommand\BibTeX{{%
    \normalfont B\kern-0.5em{\scshape i\kern-0.25em b}\kern-0.8em\TeX}}}

\setcopyright{acmcopyright}
\copyrightyear{2023}
\acmYear{2023}
\acmDOI{XXXXXXX.XXXXXXX}

\acmConference[ACM MM '23]{ACM Multimedia}{October 28, 2023 – November 3, 2023}{Ottawa, Canada}
\acmPrice{15.00}
\acmISBN{978-1-4503-XXXX-X/18/06}

\begin{document}

\title[Learn to Compress (LtC)]{Learn to Compress (LtC): Efficient Learning-based Streaming Video Analytics}

\author{Quazi Mishkatul Alam}
\affiliation{%
  \institution{University of California}
  \city{Riverside}
  \state{CA}
  \country{USA}}
\email{qalam001@ucr.edu}

\author{Israat Haque}
\affiliation{%
  \institution{Dalhousie University}
  \city{Halifax}
  \country{Canada}}
\email{israat@dal.ca}

\author{Nael Abu-Ghazaleh}
\affiliation{%
  \institution{University of California}
  \city{Riverside}
  \state{CA}
  \country{USA}}
\email{nael@cs.ucr.edu}

\renewcommand{\shortauthors}{Quazi et al.}

\begin{abstract}
Video analytics are often performed as cloud services in edge settings, mainly to offload computation, and also in situations where the results are not directly consumed at the video sensors. Sending high-quality video data from the edge devices can be expensive both in terms of bandwidth and power use. In order to build a streaming video analytics pipeline that makes efficient use of these resources, it is therefore imperative to reduce the size of the video stream. Traditional video compression algorithms are unaware of the semantics of the video, and can be both inefficient and harmful for the analytics performance. In this paper, we introduce \emph{LtC}, a collaborative framework between the video source and the analytics server, that efficiently learns to reduce the video streams within an analytics pipeline. Specifically, LtC uses the full-fledged analytics algorithm at the server as a teacher to train a lightweight student neural network, which is then deployed at the video source. The student network is trained to comprehend the semantic significance of various regions within the videos, which is used to differentially preserve the crucial regions in high quality while the remaining regions undergo aggressive compression. Furthermore, LtC also incorporates a novel temporal filtering algorithm based on feature-differencing to omit transmitting frames that do not contribute new information. Overall, LtC is able to use 28-35\% less bandwidth and has up to 45\% shorter response delay compared to recently published state of the art streaming frameworks while achieving similar analytics performance.
\end{abstract}





\maketitle

\section{Introduction}
Video analytics are becoming increasingly important at the edge with the widespread integration of camera sensors in many scenarios, such as autonomous systems and IoT devices, to benefit from the visual awareness of the surroundings. For example, authorities are ramping up the deployment of surveillance systems to assist in law enforcement and crime prevention \cite{one_billion_cameras}; transportation systems are moving towards partial or full automation \cite{hancock2019future}; and unmanned aerial vehicles are starting to revolutionize a number of industries and remote military operations \cite{drone_and_future}. 

In such systems, the videos captured by the cameras are often transferred to remote cloud servers for analytics
\cite{chen2015glimpse, chin2019adascale, wang2019bridging, zhang2018awstream, du2020server, li2020reducto}, but this incurs expensive video transportation. Several considerations drive this model including: (1) the analytics algorithm may require more resources than available on the end devices; or (2) the analytics consumers may not be located on the device itself ~\cite{qiu-18} or may require access to multiple streams of data ~\cite{kolar2016ctcv}. 
Transporting high-quality video to support accurate analytics, consumes significant bandwidth. 
Thus, a key research question regarding streaming video analytics is: \emph{how to reduce the volume of the video data without sacrificing accuracy with respect to the downstream analytics?}    

Conventional video compression algorithms, such as MPEG~\cite{grois2013performance} prioritize maximizing the perceived visual quality of the video over preserving the essential features for analytics~\cite{pakha2018reinventing, du2020server}. For example, these algorithms degrade the video quality uniformly across both regions that are salient for analytics (e.g., object regions) and those that are not (e.g., background regions). As a result, we are left with the unsatisfying choice of either losing accuracy by degrading the quality of the object regions along with the background regions, or losing bandwidth by keeping everything in sufficient quality to preserve the analytics accuracy~\cite{pakha2018reinventing}.

An alternative approach is semantic compression: such algorithms preserve necessary features while aggressively degrading unimportant regions, achieving high compression while preserving analytics accuracy.  A challenge in this approach is how to identify what regions are salient without running the full analytics at the source. 
Our insight is that differentiating the relevant regions from the background regions in a video is easier than the full analytics task.  In the context of object detection and classification, identifying which areas may have objects is much simpler than the full classification problem.  
Another observation that simplifies the problem is that, given a specific video context, we typically encounter a small subset of the possible objects -- for example, in a highway surveillance context, we expect different types of vehicles, but not animals, or furniture.  

LtC leverages these two observations by using a light-weight student neural network at the video source, which is trained by a Deep Neural Network (DNN) based teacher network at the server. The student network identifies the salient regions within video frames, allowing the source to differentially compress the video before sending over the network. This training method is suitable for the analytics tasks where the teacher network infers the locations of the object regions in addition to the primary task, such as, object detection and segmentation.  
We also contribute a novel temporal filtering algorithm that uses a feature based discriminator to identify successive video frames where semantic information has not changed significantly, allowing us to avoid their transmission entirely.

Since LtC operates by specializing the student network to the encountered scenarios, it is important to detect when these scenarios have changed, as it is likely to be the case in dynamic environments or with mobile cameras.  LtC also reacts to this occurrence of \emph{concept drift} when the server discovers that the information relayed by the source is not effective (e.g., the higher resolution regions do not yield objects).  We observe that the layers of the student network closer to the input (which we collectively call \emph{the encoder}) are likely to remain unchanged with the scenario since they embed general features (i.e., shapes, patterns, etc.). Instead, we train only the layers closer to the output (which we call \emph{the extension}) mapping features to posteriors, which significantly reduces the update overhead.  The student network consists of merely 0.5\% of the teacher network's total parameters, and in the event of concept drift we only need to update as little as 2.2\% of the student's parameters.

In order to ensure fair comparison, we have implemented LtC and other baselines by extending an open-source emulator ~\cite{du2020server}. The video dataset was obtained from live-streaming public surveillance cameras and videos from public streaming platforms. Additionally, we regulated the end-to-end traffic conditions using an open-source network emulator ~\cite{netem}.  When compared to the closest spatial compression baseline (DDS from SIGCOMM 2020)~\cite{du2020server}, LtC uses 23\% less bandwidth and has 21-45\% shorter response time. Moreover, it is able to filter 8-14\% more frames than the closes temporal compression baseline (Reducto, also from SIGCOMM 2020) \cite{li2020reducto}, and as a result, reduces bandwidth comsumption by 8\%. Combining both spatial and temporal compression, LtC uses 28-35\% less bandwidth while having 14-45\% shorter response time than these state-of-the-art alternatives. Also, LtC substantially outpuerforms commercial video compression standards including AWStream ~\cite{zhang2018awstream}, Cloudseg ~\cite{wang2019bridging} and Glimpse ~\cite{chen2015glimpse} both in terms of F1-score and bandwidth use.

In summary, the contributions of this paper are as follows:
\begin{itemize}[leftmargin=10pt]
    \item \textbf{One-shot semantic spatial compression using lightweight student network.} We use a student network to implement semantic compression without frequent feedback from the server.
    \item \textbf{Temporal filtering through efficient deep feature differencing.} We develop a new approach for filtering redundant frames by measuring changes in the feature space. 
    We show that this approach is more stable to unimportant changes in the video (e.g., due to wind, illumination, etc.) and therefore more effective than manually crafted features 
    because it uses context-specific knowledge \cite{li2020reducto}.
    \item \textbf{Efficient adaption to concept drift.} We introduce an efficient concept drift detection and update mechanism, significantly lowering the cost of updating the detector. 
\end{itemize}

\section{Motivation and Opportunity}
\begin{figure}[t]
\begin{subfigure}[t]{\linewidth}
\centering
\includegraphics[width=0.3\linewidth]{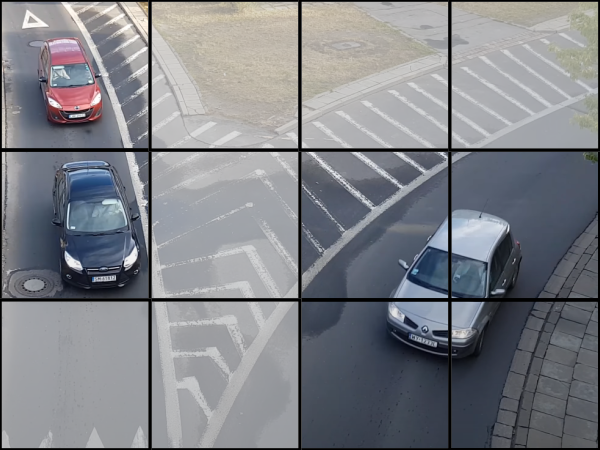} 
\caption{ Spatial compression}
\end{subfigure}
\begin{subfigure}[t]{\linewidth}
\vspace{5pt}
\centering
\includegraphics[width=0.9\linewidth]{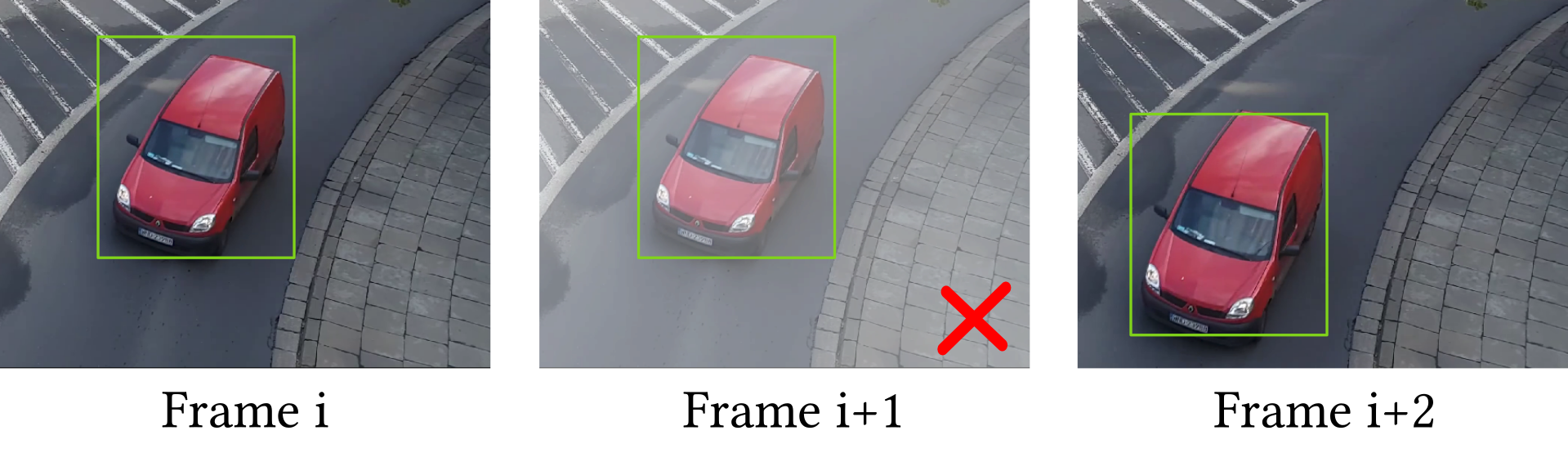}
\caption{Temporal compression}
\end{subfigure}
\caption{
Spatial compression is intra-frame, while temporal compression is inter-frame}
\label{fig:spatial_temporal}
\end{figure}

The pertinent question that arises is, \emph{why do we need different compression algorithms for video analytics?}   Conventional video compression algorithms (e.g., HVEC or MPEG) are designed to maintain human-perceived visual quality; these compression algorithms uniformly degrade the quality across the video. In contrast, the goal of semantic compression is to maintain the accuracy of downstream analytics, and therefore, by using differential compression techniques it is able to achieve a better bandwidth-accuracy tradeoffs. The process of spatial and temporal compression with respect to object classification is presented in Figure \ref{fig:spatial_temporal}.

To demonstrate the potential of semantic compression in comparison to conventional video compression in the context of video analytics, we carry out a series of experiments. Our results indicate that semantic compression offers better bandwidth-accuracy tradeoffs than conventional codecs, such as MPEG. To evaluate the performance of these compression algorithms, we calculate their F1-scores by comparing the outputs of the full-sized DNN on both the compressed and the original video data. Alternatively, this approach can be thought of as the full-sized DNN being placed at the source to act as an optimal semantic compression, which provides perfect estimates of the positions and sizes of the required regions for analytics (by the same DNN) at the server. This optimal algorithm, however, is not feasible in resource-constrained source devices, and serves only to illustrate the size of the opportunity.

\begin{figure}[t]
\centering
\includegraphics[width=0.60\linewidth]{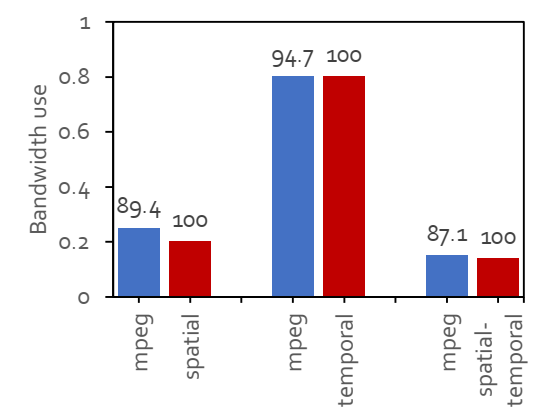} 
\caption{F1-score of optimal semantic compression vs. MPEG}
\label{fig:mpeg_comparison}
\end{figure}

  Figure \ref{fig:mpeg_comparison} presents the performance of MPEG against the optimal versions of spatial, temporal, and spatial-temporal algorithms. Each MPEG bar is configured such that it consumes either similar or higher bandwidth than the semantic compression. In all cases, the MPEG suffers a 5-13\% F1-score drop for similar bandwidth use. This is a substantial loss in terms of F1-score: over the past 5 years, the improvement in the state of the art performance in ImageNet classification is less than 10\% \cite{beyer2020we, stock2018convnets}. Also, to reach a 95\% F1-score, MPEG is able to compress only 20\% of the highest quality video (e.g., the MPEG bar with 94.7\% F1-score), while spatio-temporal compression (albeit under optimal conditions) is able to compress over 80\%.

\begin{figure}[t]
\centering
\begin{subfigure}{.49\linewidth}
    \centering
    \includegraphics[width=0.95\linewidth]{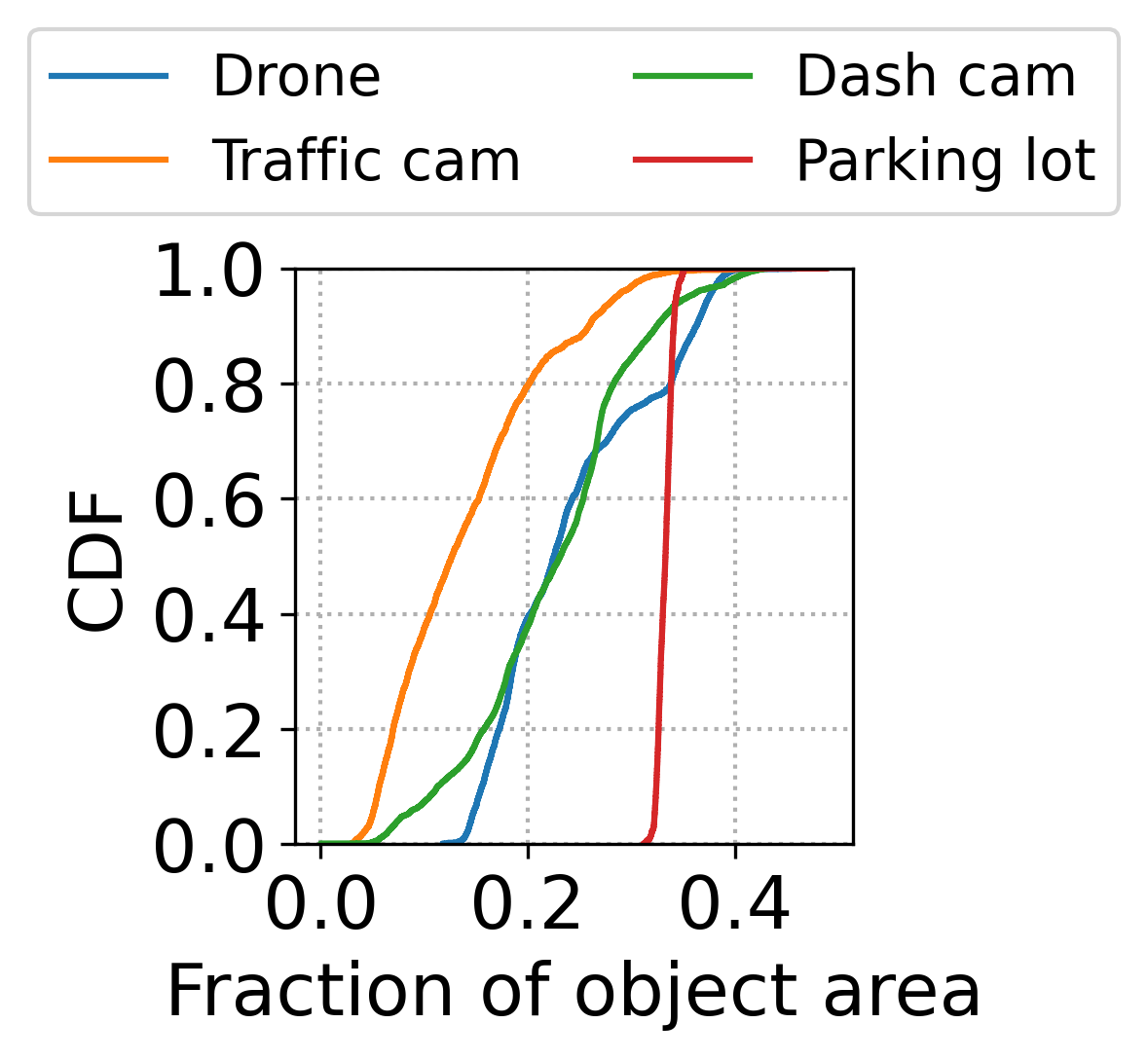}
    \caption{Useful frame-regions}
    \label{fig:spatial_cdf}
\end{subfigure}
\begin{subfigure}{.49\linewidth}
    \centering
    \includegraphics[width=0.95\linewidth]{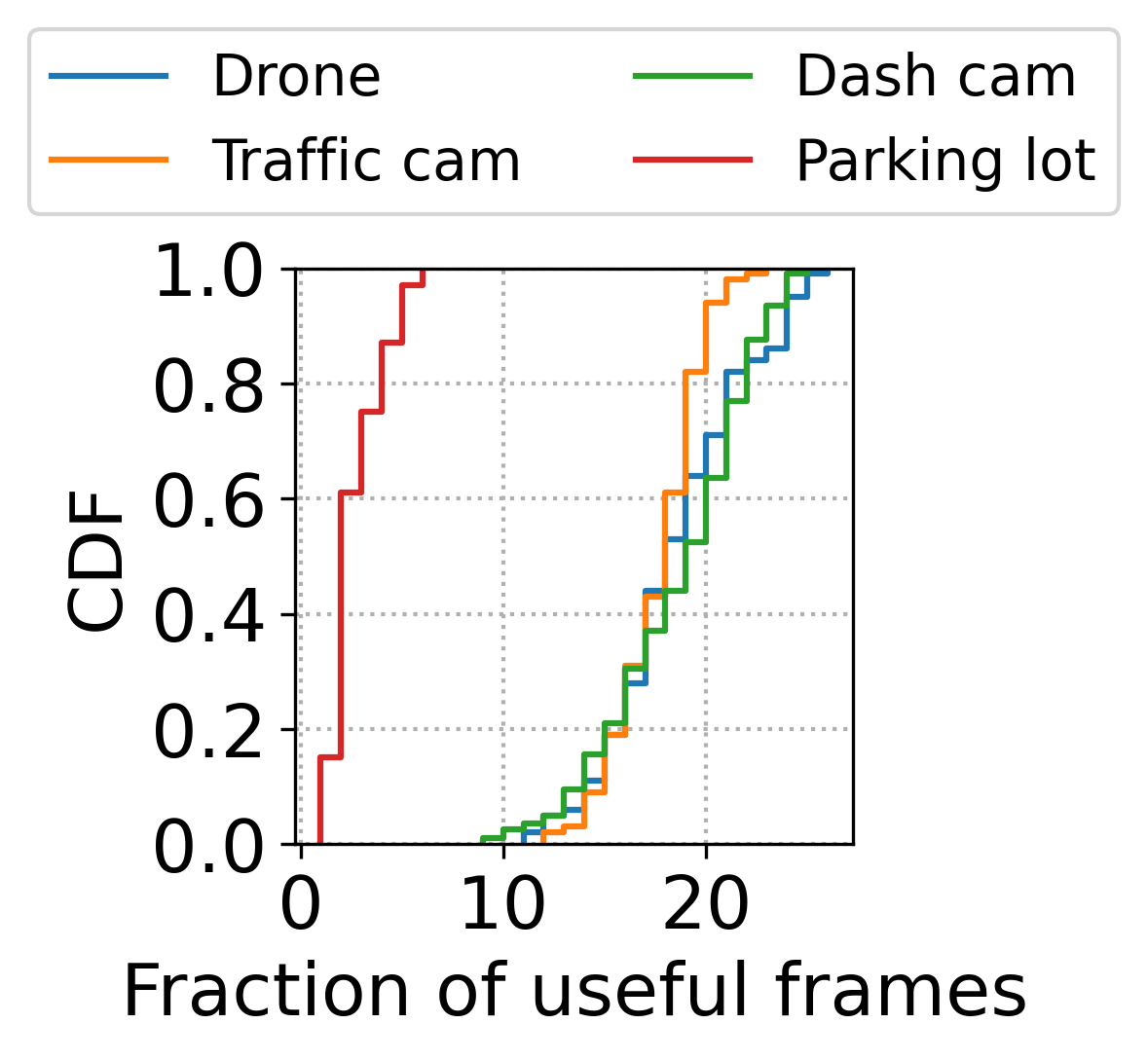}
    \caption{Useful frames}
    \label{fig:temporal_cdf}
\end{subfigure}
\caption{Spatial and temporal redundancies in videos}
\end{figure}

We also use the optimal semantic compression algorithm to characterize the redundancies present along the spatial and temporal dimensions in real videos from our dataset listed in Table \ref{tab:dataset}. Generally, in these videos the object regions constitute a small fraction of the size of the field of view. Across all the videos, around 80\% of the video frames have 33\% or fewer object regions (Figure \ref{fig:spatial_cdf}). Also, approximately in 80\% of the 1-second batches sampled from these videos, only 20 frames in 30 frames-per-second (FPS) videos and 5 frames in 15 FPS videos are necessary for analytics (Figure \ref{fig:temporal_cdf}).

\section{LtC Design and Implementation}

\begin{figure}[t]
    \centering
    \hspace{-45pt}
    \includegraphics[width=0.9\linewidth]{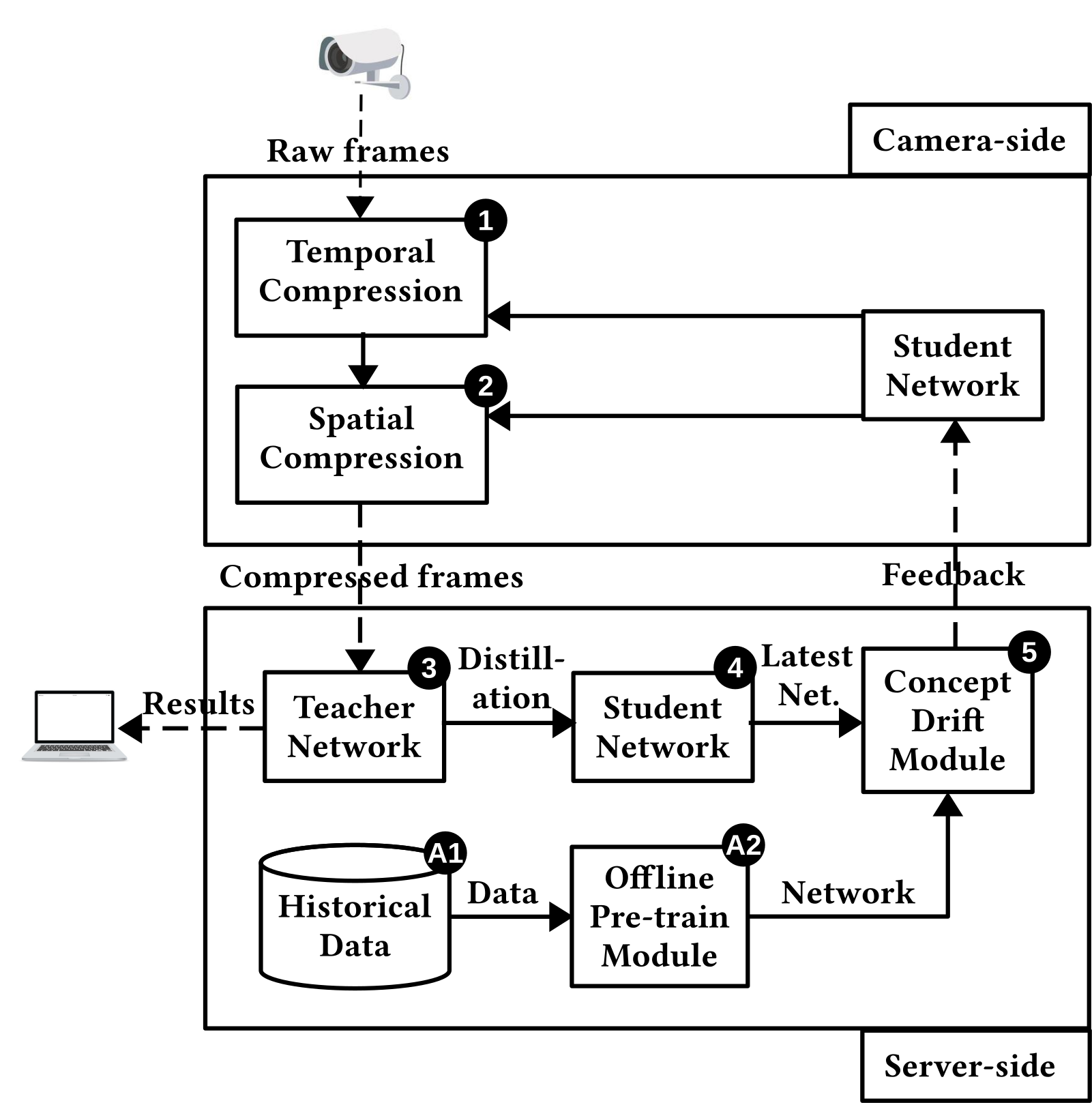}
    \caption{Overall architecture of LtC}
    \label{fig:architecture}
\end{figure}

The overall architecture of LtC is presented in Figure ~\ref{fig:architecture}. 
The source-side component is centered around a compact student neural network, which is able to identify the object regions in order to perform spatial compression. The student network also provides a tensor based feature vector extracted from its intermediate layers to be used for temporal filtering. As the camera captures the video, the source accumulates a fixed-size batch of frames to process at a time. 
The student network applies both spatial and temporal compression before it sends the batch over to the server. As the batch arrives at the server, the teacher network performs analytics and looks for concept drift. Notably, the student network is pretrained on historical data allowing itself to quickly adapt to the present scenario with only a few frames.

\subsection{The student-teacher framework}

The student-teacher framework is at the heart of LtC's system design. This method of training enables a large neural network to distill its knowledge into a smaller network \cite{hinton2015distilling, wang2021knowledge}. As the student network specializes only on a subset of the whole data, and tackles a much simpler task than the actual analytics, its size can be kept small. Moreover, initial pretraining and subsequent updating allows for fast retraining with only a small amount of new data.



In the context of LtC, the teacher DNN trains the student network to determine the objectness (likelihood of an object being present) in the regions within a frame.   In our approach, the video frames are divided into an array of non-overlapping equal-size regions. These regions are labeled using the results of the object detection from the teacher network $\mathcal{T}$, to be used in the training of the student network $\mathcal{S}$. 

The training starts as the camera sends $\mathcal{N}$ video frames $F^\mathcal{I}=[f^\mathcal{I}_1, f^\mathcal{I}_2, ..., f^\mathcal{I}_{\mathcal{N}}]$ to the server in $\mathcal{I}$th iteration. 
Upon receiving the frames, the teacher network $\mathcal{T}$ generates $C_f$ bounding boxes $BB^f=[bb^f_1, bb^f_2, ..., bb^f_{C_f}]$ around objects in frame $f$. 
Each frame $f$ is split into a $\mathcal{L}\times\mathcal{L}$ array $X^f = [..., x^f_{ij}, ...]$ of non-overlapping same-sized regions, where $i$ and $j$ are 2D indices of a region in the frame. We denote the posterior from the teacher $\mathcal{T}$ and the student $\mathcal{S}$ networks as $P_\mathcal{T}(y | x^f_{ij})$ and $P_\mathcal{S}(y | x^f_{ij})$, respectively, where $y$ is the measure of objectness. We use \textit{Intersection over Union (IoU)} function to determine $P_\mathcal{T}(y | x^f_{ij})$ in the following manner:
\begin{equation}
    \mathcal{S}(x^f_{ij}) = P_\mathcal{T}(y | x^f_{ij}) = \begin{cases}
			1, & \max^{C_f}_{k = 1} IoU(x^f_{ij}, bb^f_k) > 0.5\\
            0, & \text{otherwise}
		 \end{cases}    
\end{equation}

In order to train a student network that approximates a pretrained teacher network whose parameters are frozen, we only have to minimize the Kullback–Leibler (KL)-divergence between the two distributions with respect to the parameters of the student network, which is equivalent to minimizing the following loss function in each iteration $\mathcal{I}$:
\begin{equation}
    \mathcal{L}(\theta_S) = -\sum^{\mathcal{N}}_{f = 1 } \sum^{\mathcal{L}}_{i = 1}\sum^{\mathcal{L}}_{j = 1} P_\mathcal{T}(y | x^f_{ij}) \log{P_\mathcal{S}(y | x^f_{ij};\theta_S)},
\end{equation}
where $\theta_S$ are the parameters of the student network.
\begin{table}[t]
\centering
\resizebox{0.8\linewidth}{!}{
\begin{tabular}{l|ccc} 
\hline
\multicolumn{1}{l}{}                 & Layer type         & Output Shape & Param \#  \\ 
\hline\hline
\multirow{7}{*}{\rotatebox{90}{Encoder}}   & Input Layer          & 28x28x3  & 0         \\
                                     & 2D Convolution Layer & 28x28x16 & 448      \\
                                     & 2D Max Pooling       & 14x14x16 & 0         \\
                                     & 2D Convolution Layer & 14x14x32 & 4640     \\
                                     & 2D Max Pooling       & 7x7x32   & 0         \\
                                     & Flatten Layer        & 1568         & 0         \\
                                     & Dense Layer          & 128           & 200832    \\ 
\hline\hline
\multirow{5}{*}{\rotatebox{90}{\hspace{7pt}Extension}} & Input Layer          & 128           & 0         \\
                                     & Dense Layer          & 32           & 4128      \\
                                     & Dense Layer          & 16          & 528      \\
                                     & Output Layer         & 1            & 17        \\
\hline\hline
\end{tabular}}
\caption{Architecture of the student network} 
\label{tab:student_network_layers}
\end{table}

 \begin{table}[t]
\centering
\resizebox{0.8\linewidth}{!}{
\begin{tabular}{ccc} 
\hline
Network type                                                        & Param \# & Size      \\ 
\hline
Faster R-CNN ResNet-101 (Teacher) & 44M      & 196.5 MB  \\
Encoder (Student)                                                            & 205920   & 844.2 KB    \\
Extension (Student)                                                          & 4673     & 36.5 KB     \\
Student Total                                                             & 210593   & 880.7 KB    \\
\hline
\end{tabular}}
\caption{Sizes of the teacher, and the student network (broken into the encoder, the extension)}
\vspace{-0.2in}
\label{tab:relative_sizes}
\end{table}

We use the Faster R-CNN ResNet-101 ~\cite{ren2015faster} as our teacher network. 
The student network (shown in Table \ref{tab:student_network_layers}) is inspired by the VGG-16 model~\cite{simonyan2014very}, which also uses alternating convolution and max pooling layers. 
The parameters of the full network as well as the student can be seen in Table~\ref{tab:relative_sizes}. Notably, the teacher is over 200x larger than the student network. 

The student network $\mathcal{S}$ is divided into two sub-networks: (1) Encoder $\mathcal{U}$ and (2) Extension $\mathcal{V}$ (sizes in Figure \ref{tab:relative_sizes}). The encoder is responsible for converting the input frame-region into a feature vector. The extension is the later part, which translates the feature vector into an objectness score. In literature \cite{zhang2015design, pan2009survey}, this kind of split is most commonly performed to rapidly train a neural network for a task by reusing the features of a pretrained model on a similar task as a starting point. For related tasks, the encoder part of the network remains relatively unchanged, as it captures low-level features, such as, shapes, patterns, and other features, which do not drift significantly for similar objects. High-level features captured by the extension, such as objectness, tend to be more sensitive to environment changes. In the context of compression for video analytics, this encoder-extension split is a novel contribution of LtC. This split has two benefits: (1) it provides a vector of semantic features that is built implicitly during training, which we leverage for temporal filtering, and (2) it allows for efficient update of the student network by updating only the extension. 


\subsection{Temporal filtering using embedded features}

Temporal filtering traditionally involves performing a comparison between frames to see if the new frame includes sufficient new information.   This comparison typically uses manually crafted low-level features that operate directly on the input image (e.g., SIFT and HOG)~\cite{li2020reducto}. Such features can be sensitive to small changes to the input resulting from the presence of wind or similar effects that do not substantially change the semantic features of the image.

We propose a novel temporal filtering approach that uses directly embedding in the feature space of the student network to detect salient changes in the image.  Specifically, we use the differences in terms of the feature vector acquired from the encoder to judge whether a frame should be discarded.  
The main advantage using this feature vector is: it is highly contextual and up-to-date compared to other manually crafted features. Moreover, it is more robust against noises in the input caused by environmental factors such as illumination or weather.

For temporal filtering, we calculate the sum of the differences between feature vectors in consecutive frames. If this difference is below a threshold value, then that pair is considered identical.  
We define a difference function between frames $f_1$ and $f_2$ as follows:
\begin{equation}
    D(f_1, f_2) = \sum^{\mathcal{L}}_{i = 1}\sum^{\mathcal{L}}_{j = 1} {\lVert \mathcal{U}(x^{f_1}_{ij}) - \mathcal{U}(x^{f_2}_{ij}) \rVert}^2_2
\end{equation}

Instead of sending $\mathcal{N}$ frames $F^\mathcal{I}$ = $[f^\mathcal{I}_1, f^\mathcal{I}_2, ..., f^\mathcal{I}_{\mathcal{N}}]$ the camera uses the difference function to divide $F^\mathcal{I}$ into $\mathcal{M}$ partitions $[p^\mathcal{I}_1, p^\mathcal{I}_2, ..., p^\mathcal{I}_{\mathcal{M}}]$ of consecutive frames in the $\mathcal{I}$th iteration, where $1 \leq \mathcal{M} \leq \mathcal{N}$ and all frames in a partition $p$ satisfy: 
\begin{equation}
    \max D(f_1, f_2) < th; \forall f_1, f_2 \in p 
\end{equation}

Lastly, we construct the list of filtered frames $\hat{F}^\mathcal{I}$ by selecting a representative frame $f$ from a partition $p$ as follows:
\begin{equation}
    f = \argmin_{x \in p} \sum_{\substack{x \neq y \\ y \in p}} D(x, y)
\end{equation}

\subsection{Spatial compression using objectness score}

Spatial compression is performed after the temporal filtering. Based on the objectness score of a frame-region given by the student network, that region is either preserved in high-quality or degraded. 
LtC uses the \emph{posterior} $\mathcal{V}(\mathcal{U}(x^f_{ij}))$ to define an identity function as follows:
\begin{equation}
        I(x_{ij})= \begin{cases}
            1, & \mathcal{V}(\mathcal{U}(x^f_{ij})) > 0.5\\
            0, & \text{otherwise}
        \end{cases}
\end{equation}
We keep a region $x_{ij}$ if $I(x_{ij})$ is 1, or compress it otherwise. Each frame $f$ is converted into a $\mathcal{L}\times\mathcal{L}$ array $X^f = [..., I(x_{ij})x^f_{ij}, ...]$ of non-overlapping same-sized regions, where $i$ and $j$ are 2D indices of a region in the frame.

\subsection{Updating the student network}
The student network may become less effective following a significant change in the environment or scene dynamics, leading to a concept drift. When concept drift is detected at the server, an update of the student network is triggered. A copy of the currently operating student network is maintained at the server, which is evaluated by the teacher network in each iteration to check if indeed object regions yield high objectness score. If the percentage falls below a fixed threshold, the server initiates a student-teacher training, causing the student network to change partially (freezing all other parameters but the extension) or, in rare cases, wholly. Based on the degree of change, the server then sends either the extension or the new student network to the source as an update. We observe that 85\% of the updates consist only of the extension, which is approximately 25x smaller than the encoder as can be seen in Table \ref{tab:relative_sizes}.

\subsection{Discussion and General Properties of LtC}

\begin{figure}[t]
    \centering
    \includegraphics[width=0.42\linewidth]{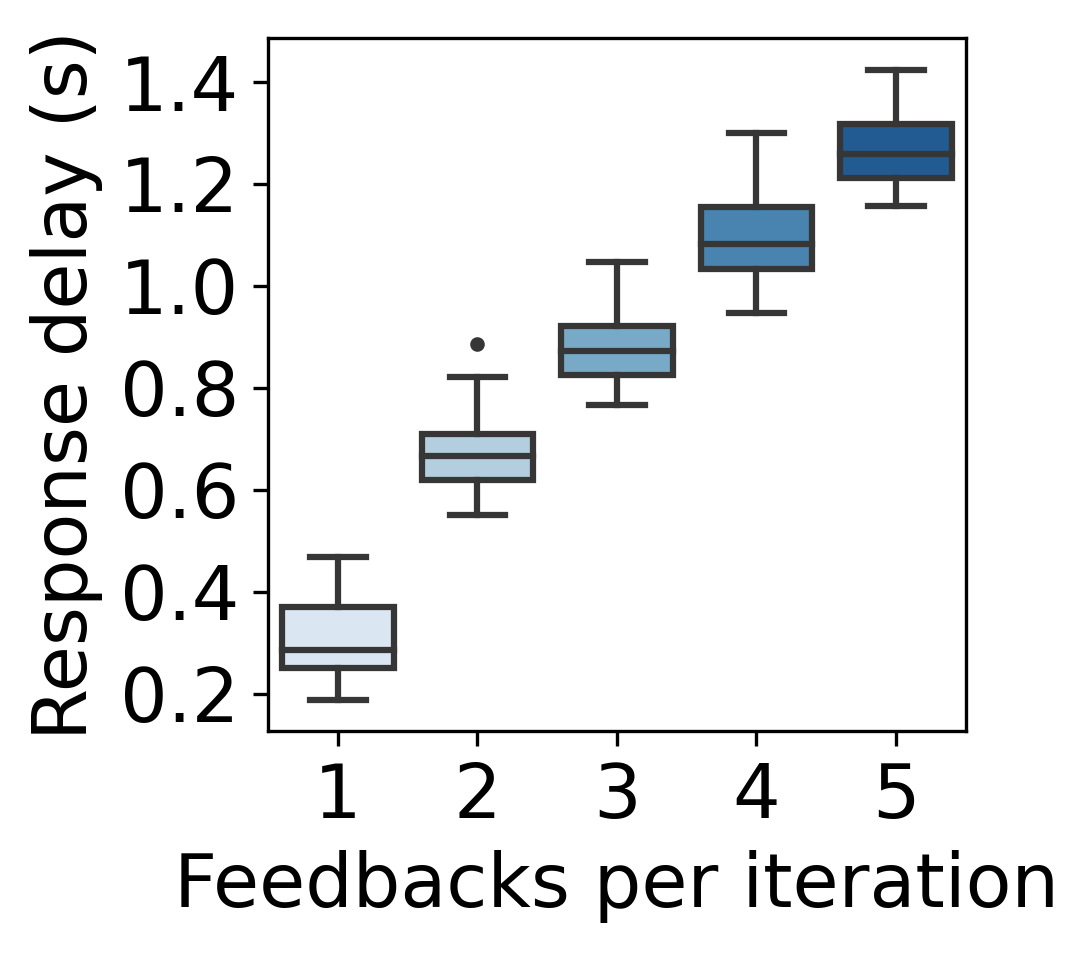}
    \caption{DDS experiences delays proportional to the number of feedback rounds used; LtC is one shot}
    \label{fig:i_dds_suboptimality}
\end{figure}

A state-of-the-art spatial compression algorithm, DDS \cite{du2020server}, uses server-side feedback in order to identify the regions that require high-quality transmission. Specifically, a low-quality baseline transmission of the video is first sent to the server, which uses the full-sized DNN to identify the regions of interest that it requires in higher resolution.  It sends a request back to the source, which in turn resends those regions at a higher quality.  
Although this approach is able to achieve reasonably high F1-score, it results in significant delay in server response. This delay includes the network delay incurred during multiple requests and responses, as well as the server processing delay after each response. Moreover, the number of feedback used in the process adversely affects the response delay. These delays are shown in Figure \ref{fig:i_dds_suboptimality} as a function of the number of feedback rounds used for each batch of frames. LtC solves this problem by training the student network to identify the regions of interest, and the compressed video is sent in a single shot.  Although the student network is smaller than the teacher network, LtC has the advantage of operating on the full resolution video; this is in contrast to DDS which transmits a low resolution video to detect the areas with objects.

\begin{figure}[t]
    \centering
    \includegraphics[width=0.8\linewidth]{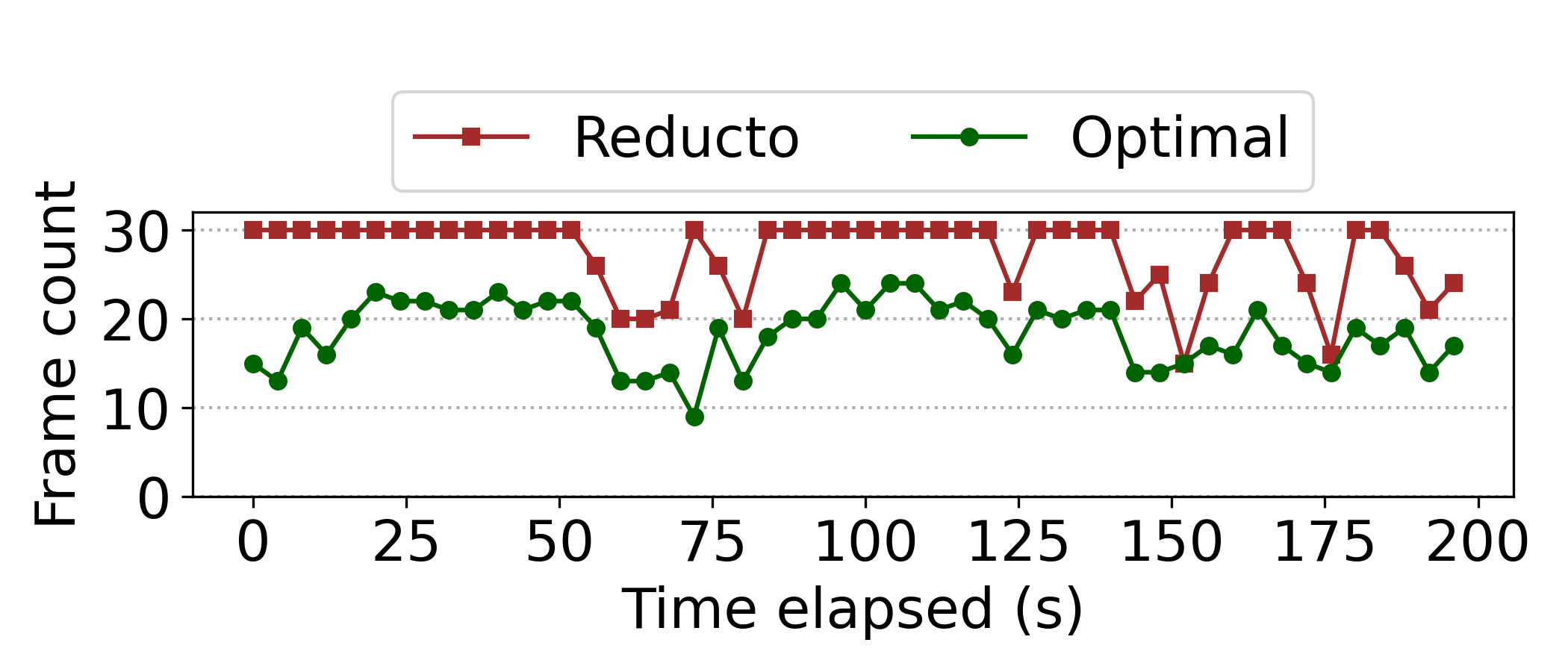}
    \caption{Reducto halts filtering during profiling phases resulting in high bandwidth use}
    \label{fig:i_reducto_suboptimality}
\end{figure}

A recently proposed temporal filtering algorithm, Reducto \cite{li2020reducto}, uses a profiling based approach, where the server profiles batches of frames against the optimal actions (what threshold to use for filtering). 
This approach is also able to achieve a high F1-score, but has high bandwidth use as the source sends unfiltered batches to the server during the profiling, as presented in Figure \ref{fig:i_reducto_suboptimality}. This profiling happens once at the beginning and every time the environment changes, and requires a large number of frames for profiling. LtC solves this problem by using the student network for profiling, which is both long-lasting due to its generalization capabilities, and is able to quickly update itself with a few new frames by transferring knowledge from the previous environments.  Additionally, Reducto does not support spatial compression.

\begin{figure}[t]
    \centering
    \includegraphics[width=0.85\linewidth]{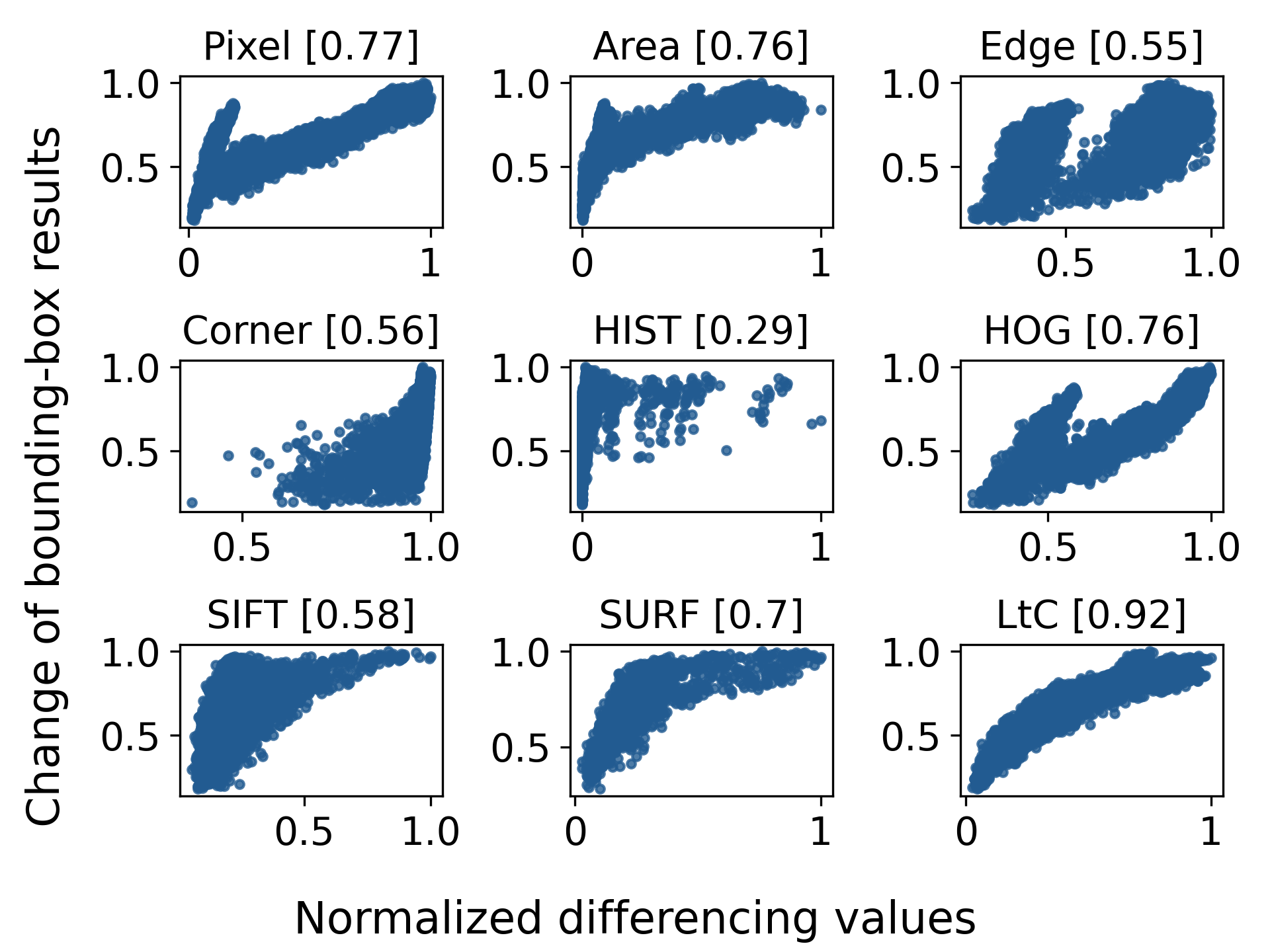}
    \caption{Correlation between the difference values and changes in the bounding-box results for different features. This experiment uses a video where the camera is mobile.}
    \label{fig:i_features}
\end{figure}

In the case of moving cameras, such as, dash cam, drone, etc., the background regions are changing constantly. As a result, use of simple features (e.g., pixel values) will result in significant difference between consecutive frames, even if the objects did not move by much. The feature vector in LtC is robust to mobility related background changes, and therefore, can recognize background regions even in the presence of mobility. In Figure \ref{fig:i_features}, we observe that for a moving camera, difference values calculated using LtC's features has the highest Pearson correlation coefficient (0.92) compared to a number of other commonly used features.

\section{Evaluation}

\noindent 
{\bf Methodology:} We used Tensorflow and Python to implement LtC within an open-source video streaming emulator for DDS~\cite{du2020server}. We extended this emulator with trace-driven emulation models for CloudSeg \cite{wang2019bridging}, Reducto \cite{li2020reducto}, and Glimpse \cite{chen2015glimpse}. This approach enables us to carry out fair comparisons within the same environment, and using similar assumptions for all the baselines. AWStream \cite{zhang2018awstream} was evaluated separately because it was not amenable to trace-driven emulation.

To the best of our knowledge LtC is the first streaming video analytics framework that uses both spatial and temporal semantic compression. While comparing with spatial (or alternatively, temporal) baselines, we turned off the temporal (and respectively, spatial) part of LtC. Consistent with prior work~\cite{du2020server, li2020reducto} we use object detection as our primary video analytics task; although it can be specialized to any analytics task that also infers the location of the object regions, such as segmentation, tagging, activity recognition, counting, etc. with alternative student training. 
We evaluate LtC using a dataset consisting of publicly available traffic surveillance videos collected from cameras deployed across various intersection of North America ~\cite{vid1, vid2, vid3, vid4, vid5, vid6, vid7, vid8}, drone videos from the VisDrone2021 dataset ~\cite{visdrone}, dashcam ~\cite{dash1, dash2} and parking lot surveillance videos from public video sharing platforms ~\cite{plot}. Details are in Table ~\ref{tab:dataset}.

\begin{table}[t]
\centering
\resizebox{0.8\linewidth}{!}{
\begin{tabular}{lllll} 
\hline
Name        & \# Videos & Total length & FPS       & Resolution    \\ 
\hline\hline
Traffic cam & 8         & live         & 15 and 30 & 4K and 1080p  \\ 
\hline
Drone       & 8         & 3 mins      & 30        & 4K and 1080p  \\ 
\hline
Dash cam    & 2         & 3 hours      & 30        & 4K            \\ 
\hline
Parking lot & 1         & 5 mins       & 15        & 1080p         \\
\hline
\end{tabular}
}
\caption{Details of the video dataset}
\vspace{-0.3in}
\label{tab:dataset}
\end{table}

We also evaluate LtC both in a resource-constrained network (1.2Mbps 100ms), and a more resource rich (100Mbps 20ms) network. We use Linux NetEm \cite{netem} to control network bandwidth and delay. 
Lastly, MPEG is used as a post-processing step to provide additional compression beyond the spatial and temporal compression, although any standard compression is equally valid for this purpose. We add this step to ensure fair comparison of LtC, against a number of baselines, such as, DDS ~\cite{du2020server}, AWStream ~\cite{zhang2018awstream}, and Cloudseg ~\cite{wang2019bridging} that integrally use MPEG in their implementations.

We evaluate LtC on the following performance metrics: 

\begin{itemize}[leftmargin=10pt]
\setlength{\itemsep}{0in}
    \item \textbf{F1-score:} We use F1-score to measure the performance of the downstream analytics. The results from using an uncompressed image is considered as the ground truth, which is consistent with the previous literature \cite{du2020server, li2020reducto}.
    \item \textbf{Bandwidth use:} The bandwidth use of transported video data is normalized with the size of the MPEG compressed image at the highest resolution (no loss in quality). 
    \item \textbf{Response delay:} This delay is the time from when a video frame is sent until the output of the analysis is ready at the server consisting of both network and processing delays. 
\end{itemize}

\subsection{Evaluating LtC's Spatial Compression}

We compare LtC with a state-of-the-art spatial compression framework DDS~\cite{du2020server}, and a few other spatial baselines, such as, AWStream \cite{zhang2018awstream} and CloudSeg \cite{wang2019bridging}. During these comparisons we disable the temporal filtering module of LtC, since the other baselines do not support temporal filtering (beyond that provided by MPEG).



\begin{figure}[t]
\centering
\begin{subfigure}{.46\linewidth}
    \centering
    \vspace{27pt}
    \includegraphics[width=\linewidth]{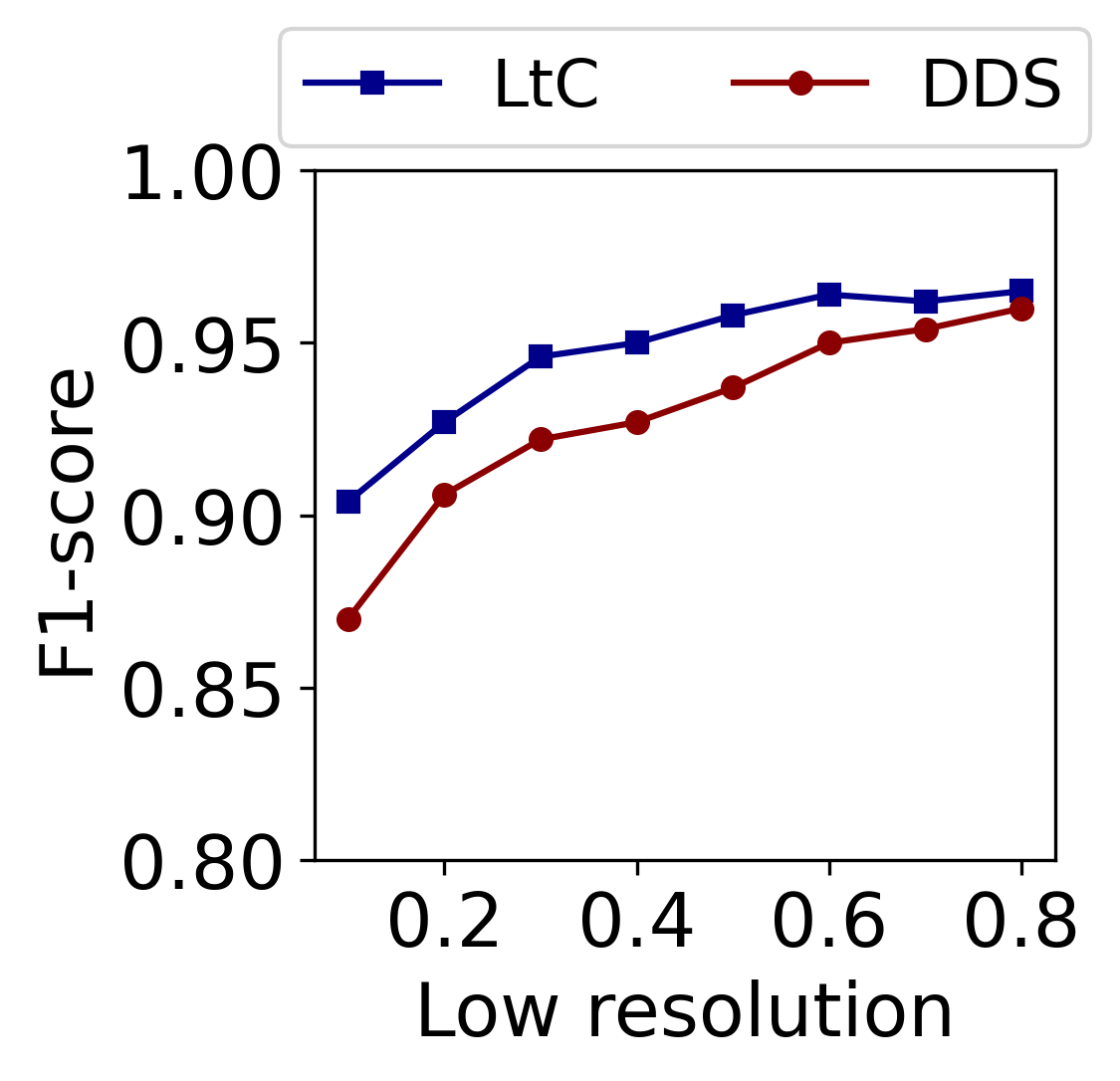}
    \caption{F1-score}
    \label{fig:accuracy_dds_ltc}
\end{subfigure}
\begin{subfigure}{.52\linewidth}
    \centering
    \includegraphics[width=\linewidth]{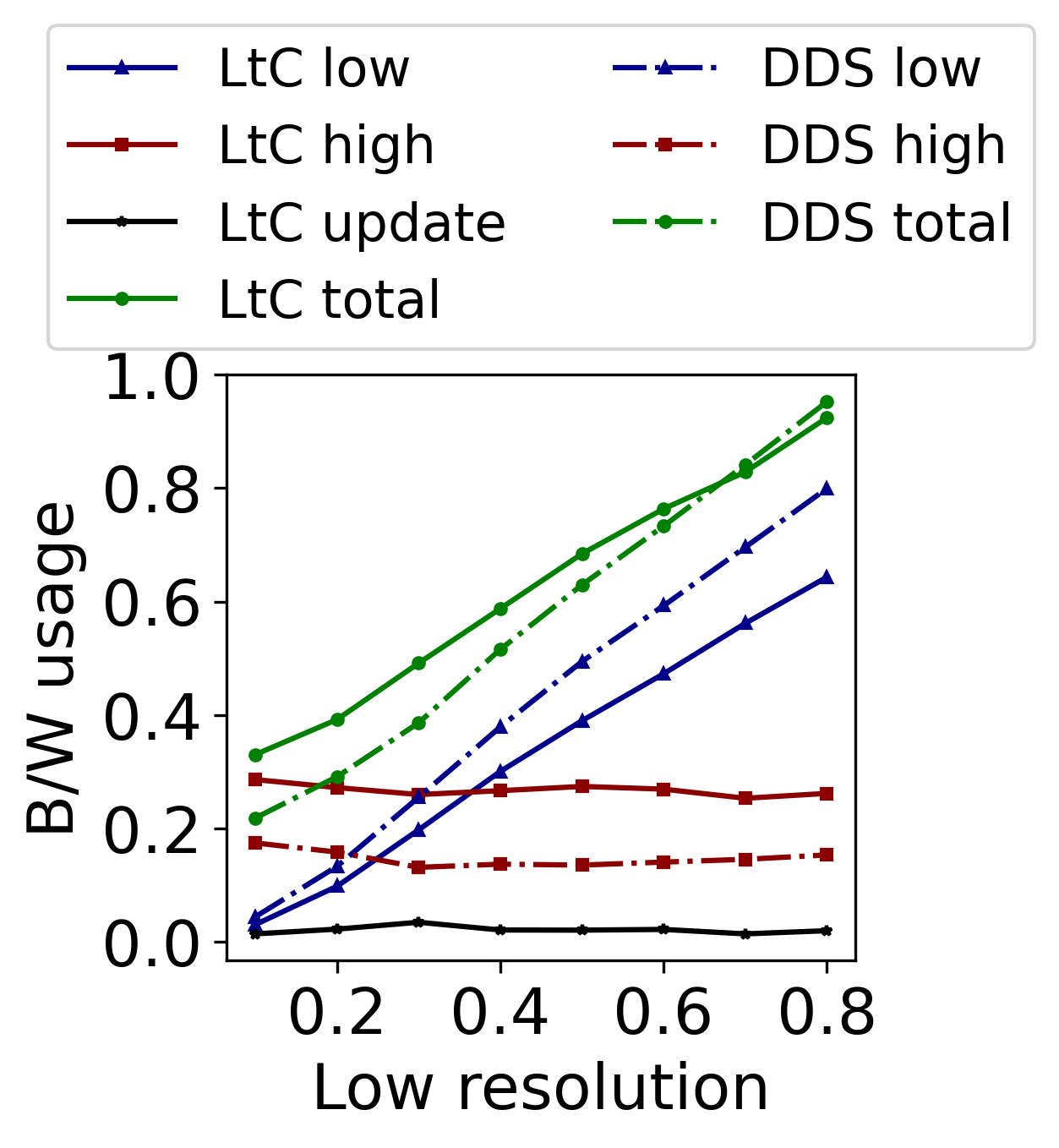}
    \caption{Bandwidth efficiency}
    \label{fig:bw_dds_ltc}
\end{subfigure}
\caption{F1-score and bandwidth efficiency of LtC and DDS against the normalized resolution of the low-quality regions. LtC consistently has higher F1-score while consuming lower total bandwidth.}
\label{fig:different_operating_points}
\end{figure}

We compare the F1-scores and total bandwidth consumption (broken down into bandwidth used for both low and high resolution phases) of LtC and DDS~\cite{du2020server} against the normalized resolutions of the low-quality regions. As presented in Figure ~\ref{fig:accuracy_dds_ltc}, LtC outperforms DDS across the entire range, specially in the low resolution range. This is because, DDS sends the low-quality regions in order to generate feedback at the server, and decreasing the resolution of these regions results in diminished feedback. Whereas, LtC has access to high-quality frames at the source, and the low-quality regions are sent to compensate for any misclassified regions.
As a result, at lower resolutions of the low-quality regions, DDS struggles to find out the locations of the high-quality regions accurately.
Also, as presented in Figure \ref{fig:bw_dds_ltc}, the sizes of the low-quality regions are less in LtC than DDS. This is because, DDS needs to send all the low-quality regions for feedback, however, LtC can exclude the low-quality regions where it sends them in high-quality, resulting in less or equivalent bandwidth use at high F1-scores.

\begin{figure}
    \centering
    \includegraphics[width=0.95\linewidth]{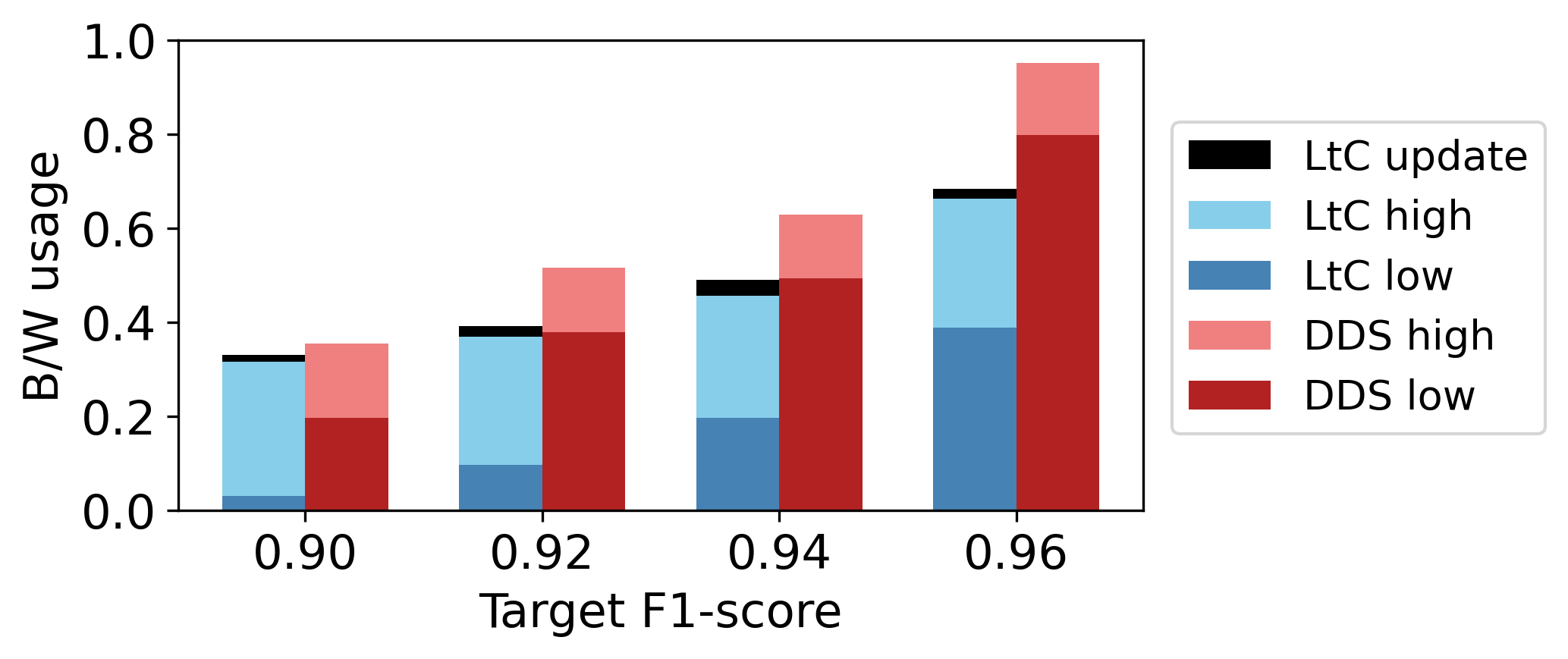}
    \caption{LtC exhibits superior bandwidth efficiency compared to DDS across a wide range of target F1-scores}
    \label{fig:accuracy_dds_ltc_alt}
\end{figure}

\noindent \textbf{Performance against target F1-scores.} For a given target F1-score, LtC can significantly outperform DDS in terms of bandwidth usage even without considering its temporal filtering. In order to generate accurate feedback regions to reach an target F1-score, DDS has to send more low quality regions (almost double) than LtC, as can be seen in Figure \ref{fig:accuracy_dds_ltc_alt}. In turn, LtC sends more high quality regions, as it has to send all the regions containing objects in high resolution. For an increasing F1-score demand, the gain of LtC over DDS also increases, as the size of the high quality regions remains the same, but the size of the low quality regions are larger and increases more rapidly for DDS. 

\noindent \textbf{Performance against different video types.} LtC and DDS perform similarly on different videos. In general, in a video with sparse frames (in terms of object presence), the bandwidth savings are higher. However, in denser videos, LtC sends comparatively more high-quality regions and uses more bandwidth. 
Therefore, LtC performs better in sparser videos, and DDS in denser videos.
\begin{figure}
    \centering
    \includegraphics[width=0.95\linewidth]{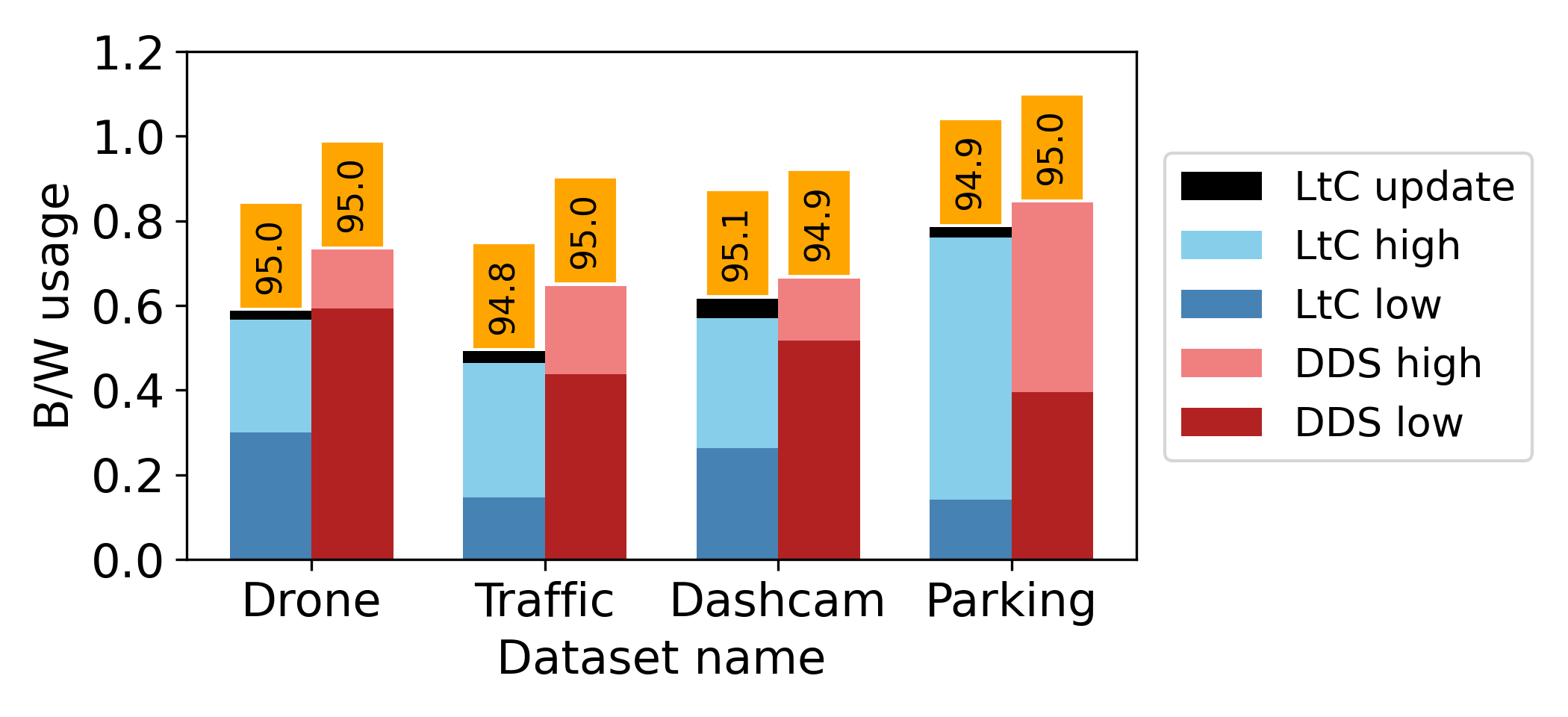}
    \caption{LtC exhibits superior bandwidth efficiency compared to DDS across different video types}
    \label{fig:bw_dds_ltc_alt}
\end{figure}

\begin{figure}[t]
\centering
\begin{subfigure}{.54\linewidth}
    \centering
    \includegraphics[width=.95\linewidth]{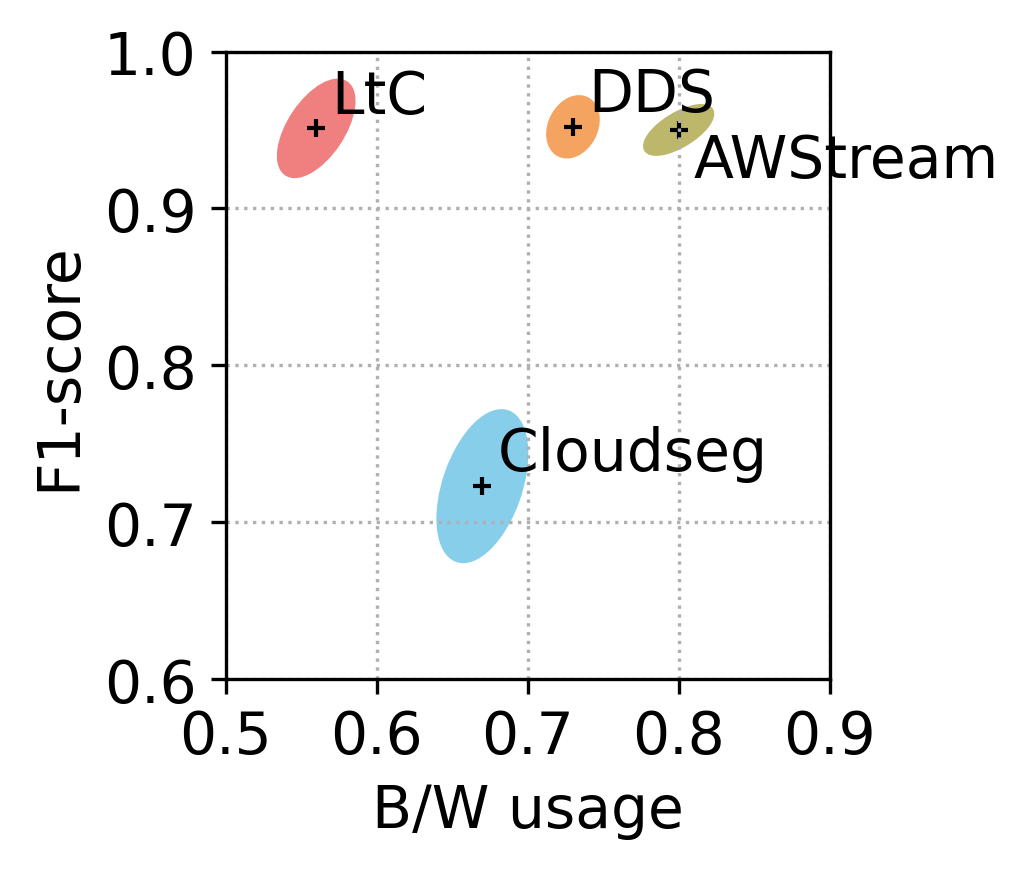}
    \caption{F1-score and bandwidth efficiency}
    \label{fig:i_spat_baselines}
\end{subfigure}
\begin{subfigure}{.45\linewidth}
    \centering
    \includegraphics[width=.95\linewidth]{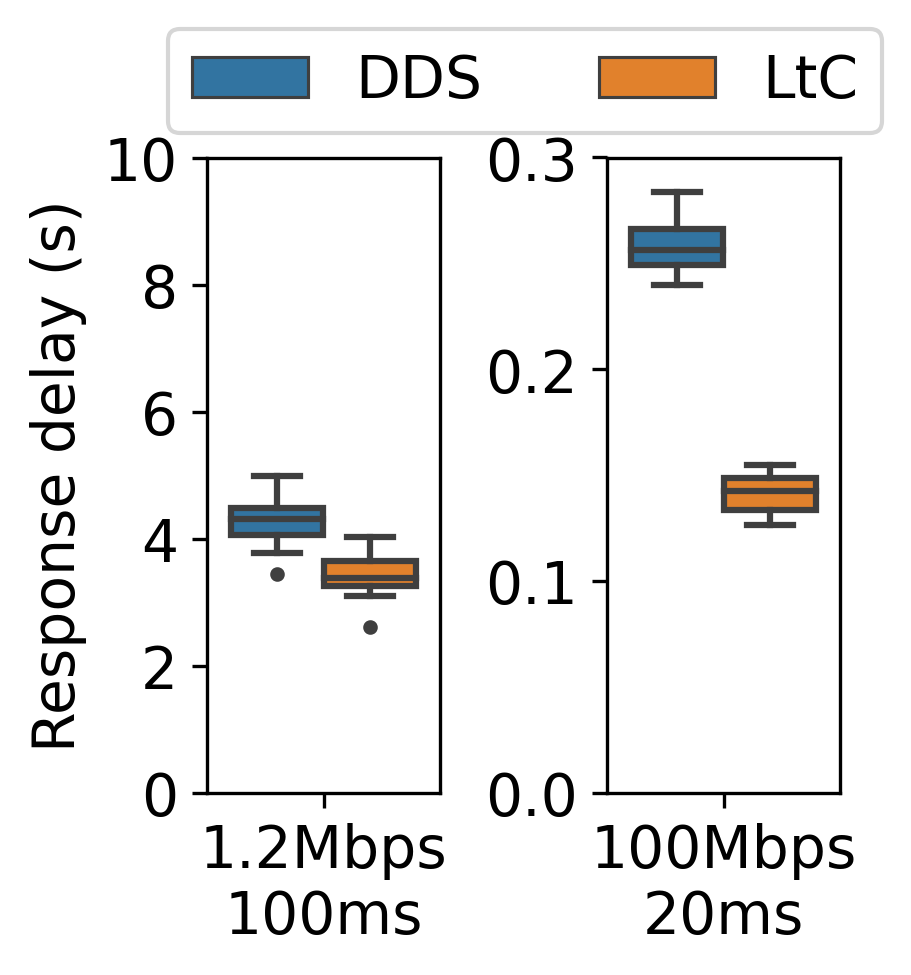}
    \caption{Response delay}
    \label{fig:i_ltc_dds_rd}
\end{subfigure}
\caption{LtC exhibits superior bandwidth-accuracy tradeoffs (upper-left is better) and lower response delay compared to other spatial compression baselines. Confidence ellipses contain 1-$\sigma$ (68\%) of results.}
\label{fig:ltc_dds_f1_bw_rd}
\end{figure}

\noindent \textbf{Performance against other spatial compression baselines.} Figure \ref{fig:i_spat_baselines} presents the F1-scores against the normalized bandwidth. Overall, LtC compresses 42\% of the highest-quality video and uses 23\% less bandwidth than DDS for the same F1-score. Other spatial compression baselines such as AWStream, and CloudSeg perform worse because they do not use semantic compression. Additionally, CloudSeg has a low F1-score because it uses downsampling and super-resolution based upsampling which appear to harm the semantic features of the video.

The response delay of LtC and DDS is presented in Figure \ref{fig:i_ltc_dds_rd}. LtC has lower network delay than DDS as it does not require feedback from the server. Also, LtC has a lower server processing delay then DDS as it only needs to process the video one time. As a result, LtC achieves an overall 21-45\% shorter response delay than DDS in our two network configurations. This advantage will likely be higher in networks with higher latency and lower bandwidth.

\subsection{Evaluating LtC's Temporal Filtering} 

In this section, we compare LtC with a state-of-the-art temporal filtering algorithm Reducto~\cite{du2020server}, and an object tracing and filtering based algorithm Glimpse \cite{chen2015glimpse}. Like before, we disable the spatial compression module if LtC to enable a fair comparison.  

\begin{figure}[t!]
    \centering
    \includegraphics[width=0.9\linewidth]{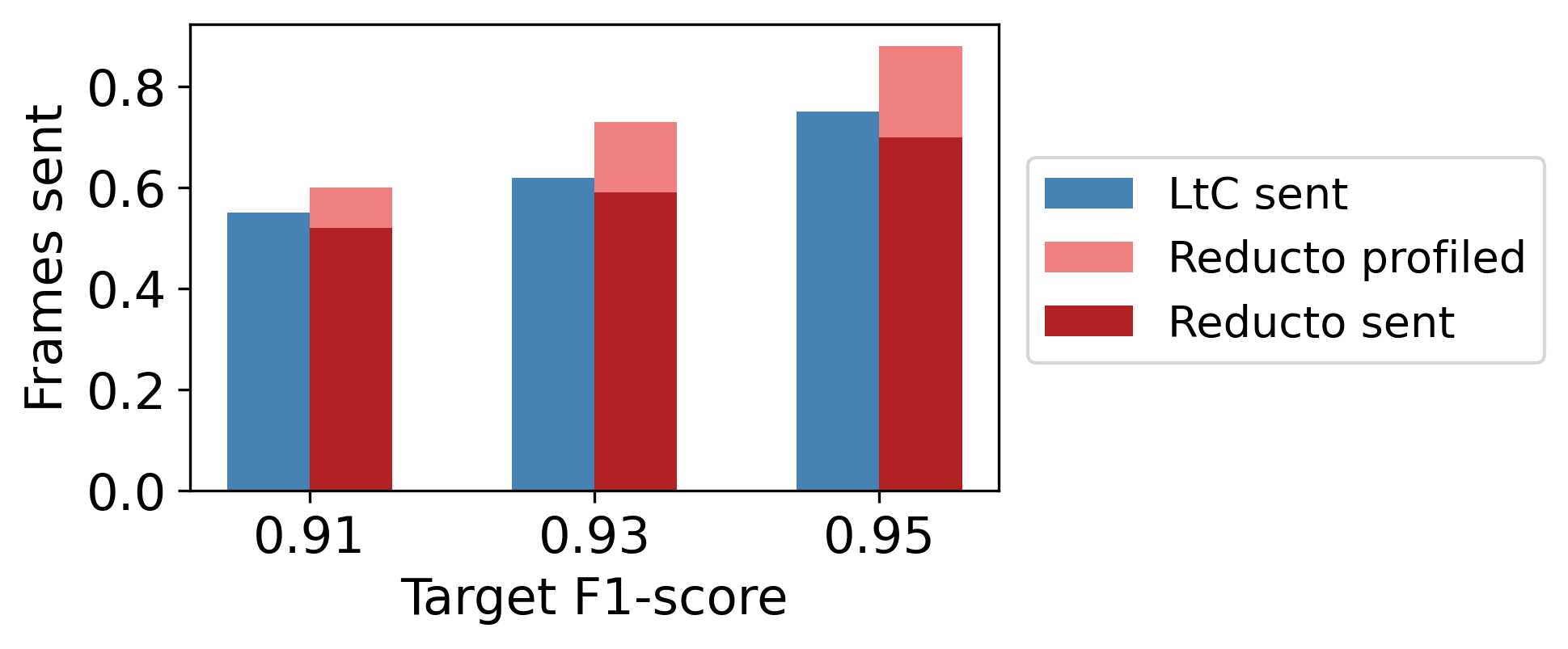}
    \caption{LtC sends out less frames compared to Reducto across a wide range of target F1-score}
    \label{fig:acc_wise_red}
\end{figure}

\noindent\textbf{Performance against target F1-score.} For a given target F1-score, LtC  outperforms Reducto in terms of number of frames filtered. In order to reach a high target F1-score, Reducto becomes conservative about using its existing profiles, and sends out more frames for profiling, as can be seen in Figure \ref{fig:acc_wise_red}. In turn, in LtC, the student network learns about object features implicitly while training. LtC also gains because the filtering based on its features is more accurate, leading to higher precision in identifying the correct frames to discard. 

\begin{figure}
    \centering
    \includegraphics[width=\linewidth]{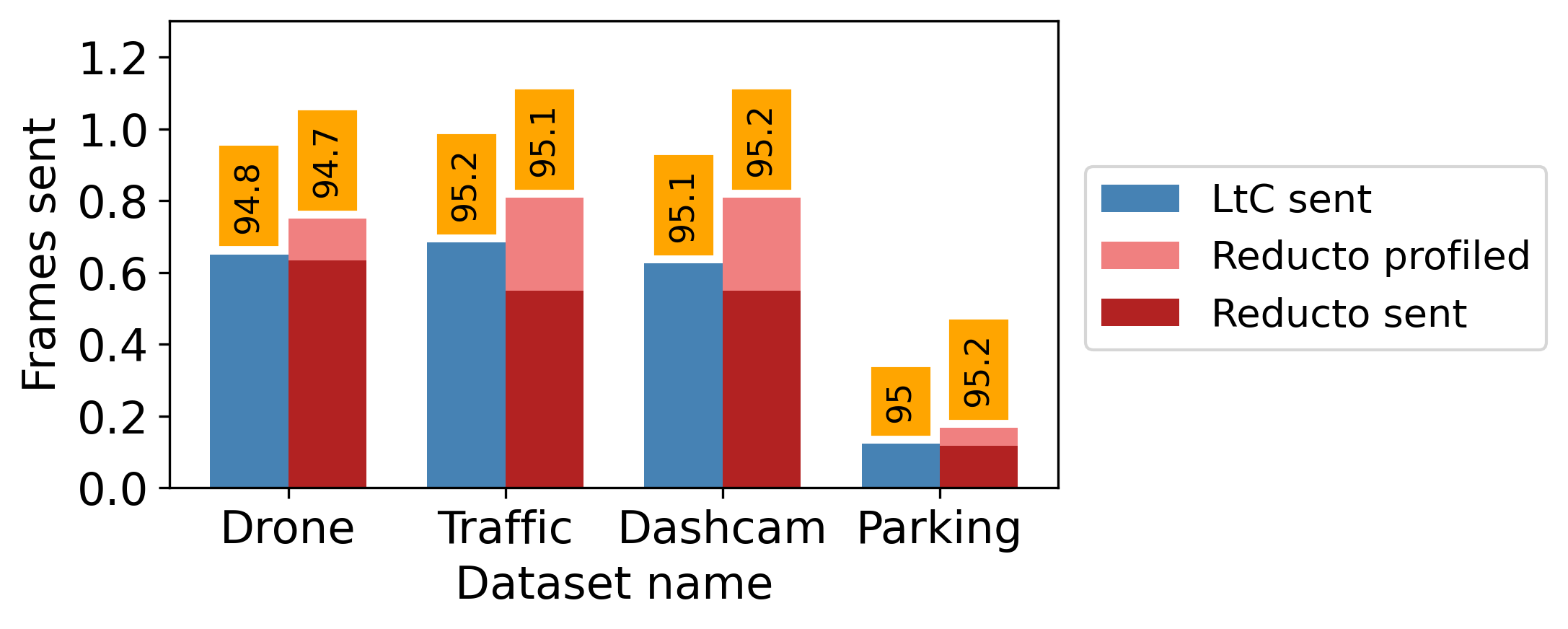}
    \caption{LtC sends out less frames compared to Reducto across different video types}
    \label{fig:vwf}
\end{figure}

\noindent\textbf{Performance against different video types.} Both LtC and Reducto are sensitive to the frame-rate and the content of the video, which determines the available opportunity for temporal filtering. Intuitively, if the frame-rate of a video is high then we have higher redundancy. In addition, there are more redundancies in a static or slowly changing video. Among the four kinds of videos in our dataset, the parking lot dataset is relatively static compared to the other videos (cars are only going in and out of the parking lot occasionally). In Figure \ref{fig:vwf}, we can see that for the parking lot videos both DDS and LtC is able to filter more frames compared to the other videos in the dataset.

\begin{figure}
    \centering
    \includegraphics[width=.5\linewidth]{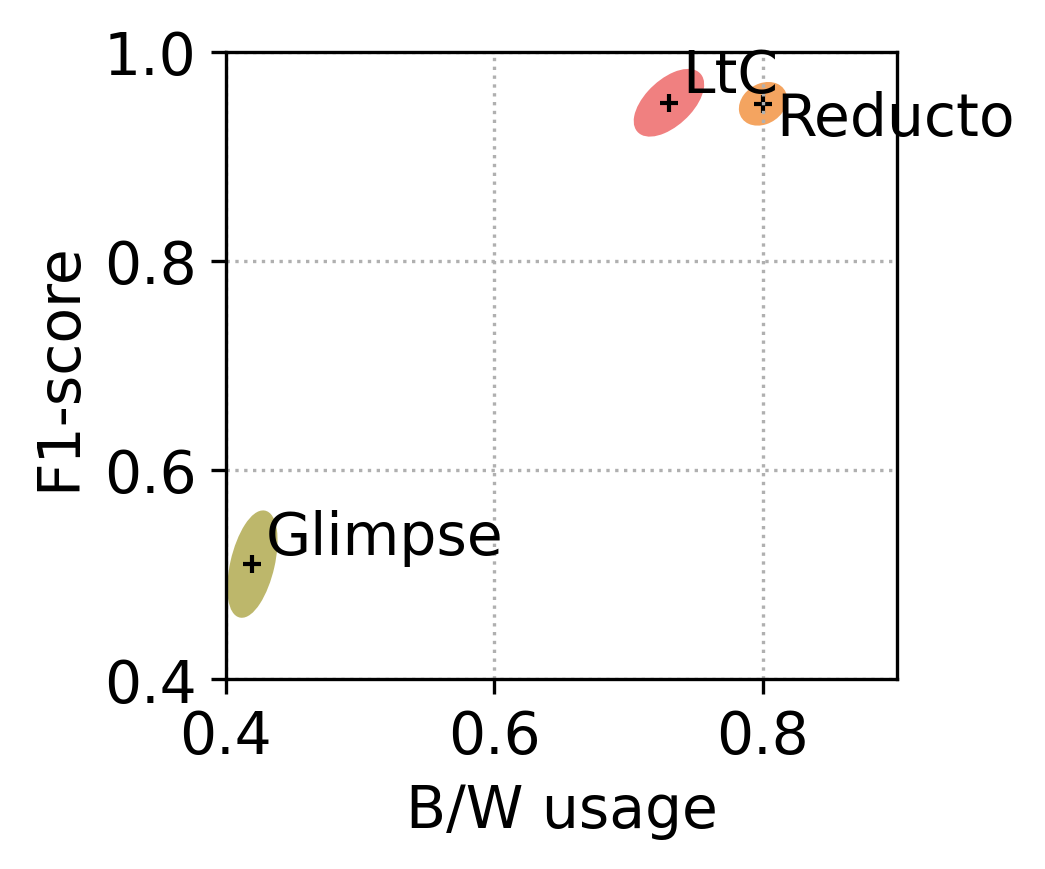}
    \caption{LtC exhibits superior bandwidth-accuracy tradeoffs compared to other temporal filtering baselines}
    \label{fig:i_tmp_baselines}
\end{figure}

\noindent\textbf{Performance against other temporal compression baselines.}  Figure \ref{fig:i_tmp_baselines} shows that LtC is able to filter 8-14\% more frames, and uses 8\% less bandwidth than Reducto for the same F1-score. The other baseline, Glimpse, falls behind in F1-score, because it filters more aggressively with a suboptimal policy, leading to a loss in analytics performance.
The response delay of LtC and Reducto is similar, since hey filter at the source.

\subsection{End to end evaluation of LtC}  
\begin{figure}[t]
\centering
\begin{subfigure}{.52\linewidth}
    \centering
    \includegraphics[width=\linewidth]{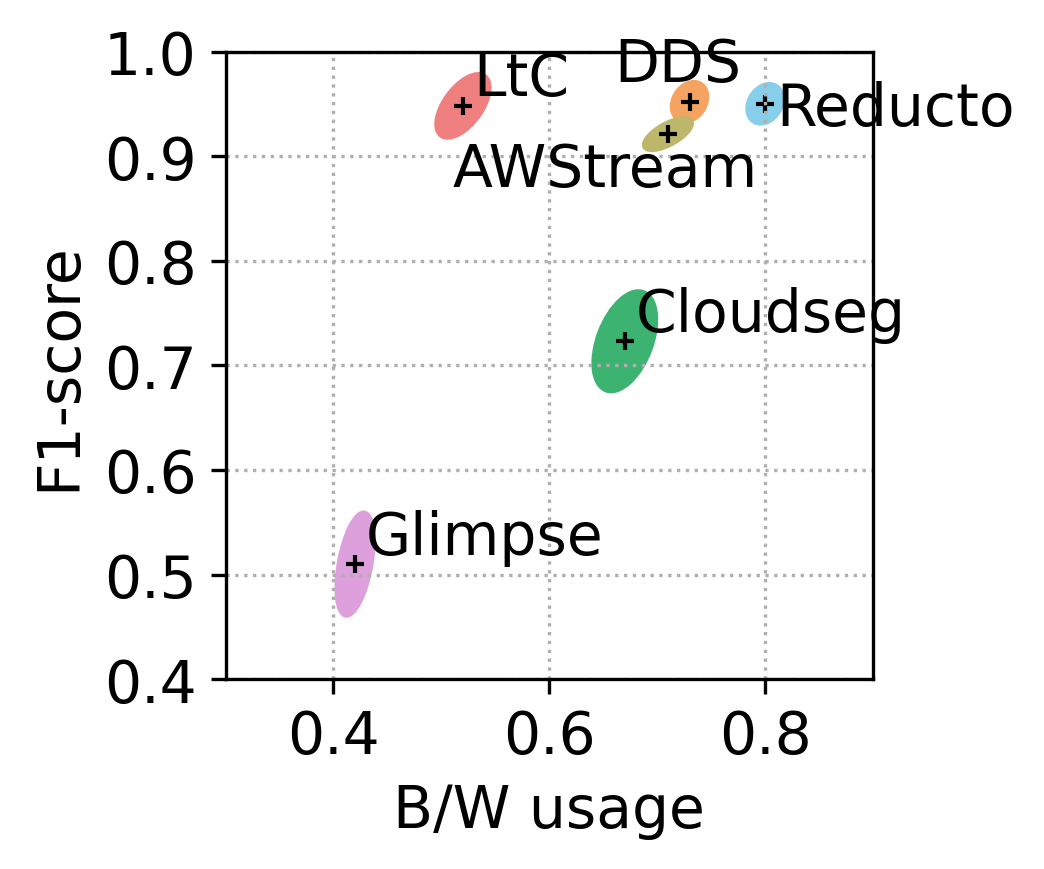}
    \caption{F1-score and  bandwidth efficiency}
    \label{fig:i_e2e_all_baselines_acc_bw}
\end{subfigure}
\begin{subfigure}{.61\linewidth}
    \centering
    \includegraphics[width=\linewidth]{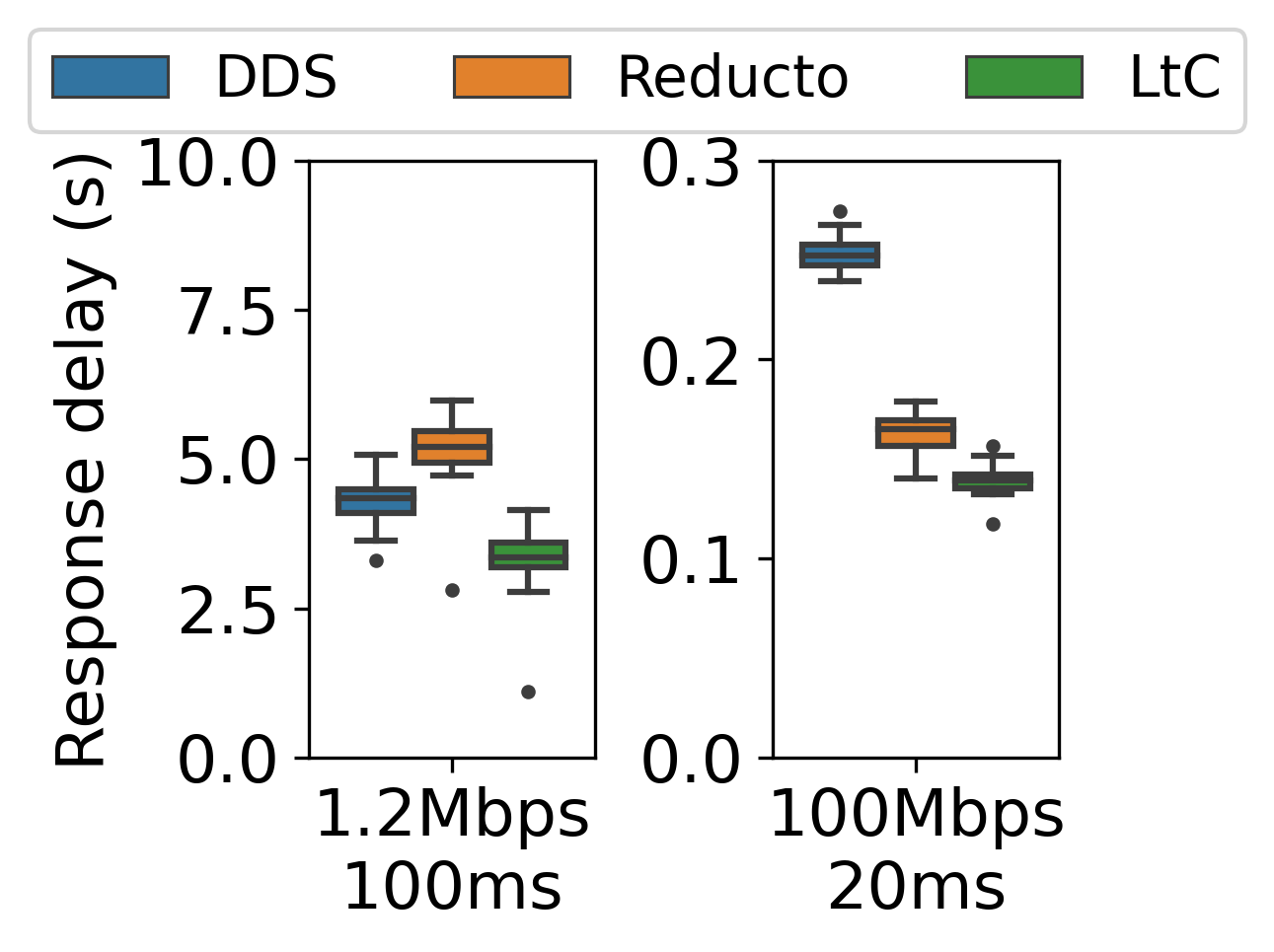}
    \caption{Response delay}
    \label{fig:i_e2e_all_baselines_rd}
\end{subfigure}
\caption{LtC exhibits superior bandwidth-accuracy tradeoffs and lower response delay compared to all other baselines in an end-to-end setting} 
\label{fig:ltc_dds_red_f1_bw_rd}
\end{figure}

In Figure \ref{fig:i_e2e_all_baselines_acc_bw}, we present the accuracy-bandwidth tradeoff of LtC against all other baselines. Overall, LtC is uses 28-35\% less bandwidth than the closest baselines DDS and Reducto. 
Moreover, for both the resource-constrained and the resource-rich network, LtC has shorter response delay than both DDS and Reducto. Specifically, LtC achieves 14-45\% shorter response delay than these baselines. Interestingly, for resource-rich network, the server processing delay of DDS exceeds the network delay, resulting in a significant advantage for LtC, as can be seen in Figure \ref{fig:i_e2e_all_baselines_rd}.

\noindent \textbf{Granularity of identifying objects.} 
The student network operates on fixed-size non-overlapping patches of each frame, to identify if they contain objects. As the number of patches along a single axis (if this number is 10, then there are 100 patches in total) increases, the performance of the student network (measured in \underline{F}rames \underline{P}er \underline{S}econd, FPS) decreases rapidly. However, increasing the number of patches results in little added benefit  beyond 20 patches. Even at a smaller number, such as 10, the bandwidth increases only slightly as the patches becomes larger. We use 16 patches along a single axis, which can run at up to 65 FPS on our machine.

\noindent \textbf{Number of layers.} The performance of LtC also varies with the number of convolution layers used in the student network. As the number increases the FPS value of the student network decreases. However, there is no added benefit when increasing beyond 2 layers (the configuration we use).
The teacher network runs at 2 FPS, while the student network runs at 65 FPS (32x faster), due to its much simpler architecture. 
\begin{figure}[t]
    \centering
    \includegraphics[width=0.9\linewidth]{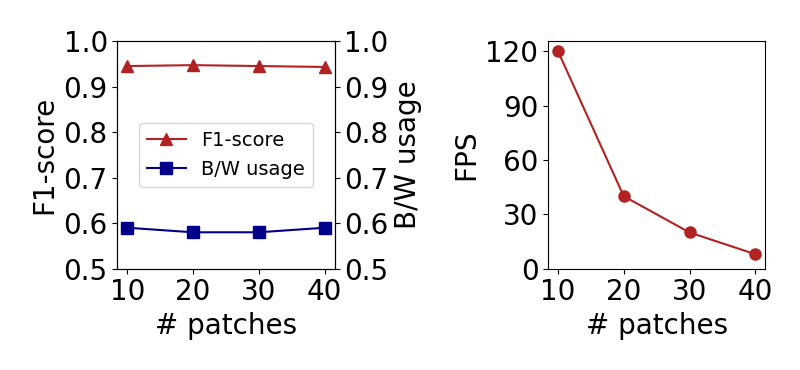}
    \caption{Reducing the number of patches has minimal effects on the bandwidth-accuracy tradeoffs, but increases FPS}
    \label{fig:patch_size}
\end{figure}
\begin{figure}[t]
    \centering
    \includegraphics[width=0.9\linewidth]{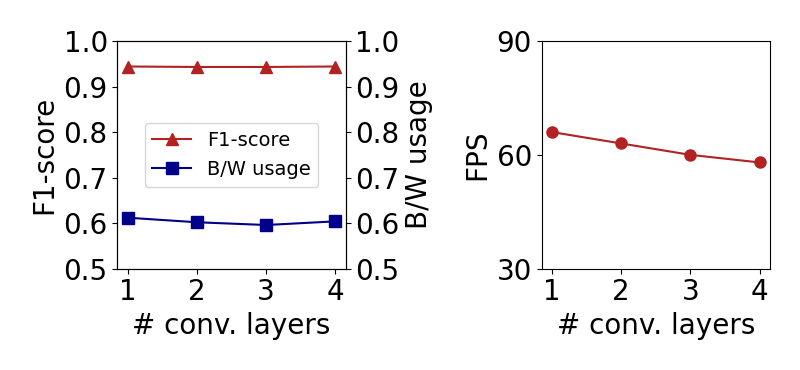}
    \caption{Reducing the number of convolutional layers has minimal effects on the bandwidth-accuracy tradeoffs, but increases FPS}
    \label{fig:model_size}
\end{figure}

\section{Related Work}


\noindent \textbf{Spatial compression.} A variety of different approaches has been developed to spatially reduce the size of video data for analytics. Preliminary designs uses source-side heuristics that are detached from the server. For example, \emph{AdaScale} \cite{chin2019adascale} and \emph{CloudSeg} \cite{wang2019bridging} reduce the outgoing video using super-resolution or other heuristics.  
Recent works tend to use server-side feedback for real-time adaptation. For example, \emph{AWStream} \cite{zhang2018awstream} collaborates with the server in order to find the optimal running configuration. The most related approach to LtC is  \emph{DDS} \cite{du2020server}, which sends high-quality patches to the initially sent low-quality image in multiple phases per iteration, based on the feedback from the server.  Split-DNNs, such as \emph{Split-brain} \cite{emmons2019cracking}, split a DNN into initial layers on the source and remaining on the server.   

\noindent \textbf{Temporal filtering.} There has been a number of systems that use source-side, or edge-based heuristics to filter frames. For example, \emph{Vigil} \cite{zhang2015design} uses a light-weight object counter model on the edge to filter out frames based on number of objects present.  \emph{FilterForward} \cite{canel2019scaling} uses light-weight application specific binary classifiers to assist in filtering. \emph{Glimpse} \cite{chen2015glimpse} uses a caching and object-tracking technique to fill out results in the omitted frames. Finally, similar to LtC, \emph{Reducto} is a feature-differencing  approach  \cite{li2020reducto}, which finds out the optimal action for a batch of frames by building a profile.  Reducto uses shallower features than LtC, and needs expensive retraining to identify optimal actions in the presence of scene dynamics.

\noindent \textbf{Tiny models.} Recently, there has been developments of tiny neural network models \cite{tiny, incremental_improv} (e.g. Tiny YOLO versions) that can run in resource-constrained end devices at high FPS, and with reasonable accuracy. However, compression using these tiny models is suboptimal as they perform poorly in videos with smaller objects.

\noindent \textbf{Optimizing data transport in sensor networks.} There is a rich body of work in the sensor network community on reducing the dimensionality of transported data (e.g., ~\cite{deshpande-04}).  Network based video streaming frameworks such as LtC and others~\cite{du2020server,li2020reducto} may be viewed as an instance of this line of work for video sensors.  In the context of video sensor networks, a number of platforms also use compression.  However, they are targeted towards deeply embedded platforms and use traditional compression~\cite{seema-11}, or even just still images, for example, using camera traps~\cite{kays-09}.

\section{Conclusion}
In this work, we present a compression algorithm that can work in real-time on video streams targeted for analytics. The proposed scheme leverages a novel student-teacher framework, and
integrates both spatial and temporal compression in a complementary way. In addition to that, LtC significantly outperforms the state of the art proposals in terms of accuracy and bandwidth use. 

\bibliographystyle{ACM-Reference-Format}
\bibliography{main}


\begin{thebibliography}{41}


\ifx \showCODEN    \undefined \def \showCODEN     #1{\unskip}     \fi
\ifx \showDOI      \undefined \def \showDOI       #1{#1}\fi
\ifx \showISBNx    \undefined \def \showISBNx     #1{\unskip}     \fi
\ifx \showISBNxiii \undefined \def \showISBNxiii  #1{\unskip}     \fi
\ifx \showISSN     \undefined \def \showISSN      #1{\unskip}     \fi
\ifx \showLCCN     \undefined \def \showLCCN      #1{\unskip}     \fi
\ifx \shownote     \undefined \def \shownote      #1{#1}          \fi
\ifx \showarticletitle \undefined \def \showarticletitle #1{#1}   \fi
\ifx \showURL      \undefined \def \showURL       {\relax}        \fi
\providecommand\bibfield[2]{#2}
\providecommand\bibinfo[2]{#2}
\providecommand\natexlab[1]{#1}
\providecommand\showeprint[2][]{arXiv:#2}

\bibitem[Adarsh et~al\mbox{.}(2020)]%
        {tiny}
\bibfield{author}{\bibinfo{person}{Pranav Adarsh}, \bibinfo{person}{Pratibha
  Rathi}, {and} \bibinfo{person}{Manoj Kumar}.}
  \bibinfo{year}{2020}\natexlab{}.
\newblock \showarticletitle{YOLO v3-Tiny: Object Detection and Recognition
  using one stage improved model}. In \bibinfo{booktitle}{\emph{2020 6th
  International Conference on Advanced Computing and Communication Systems
  (ICACCS)}}. \bibinfo{pages}{687--694}.
\newblock
\urldef\tempurl%
\url{https://doi.org/10.1109/ICACCS48705.2020.9074315}
\showDOI{\tempurl}


\bibitem[Banff Live~Cam(2021)]%
        {vid3}
\bibfield{author}{\bibinfo{person}{Canada Banff Live~Cam, Alberta}.}
  \bibinfo{year}{2021}\natexlab{}.
\newblock
\newblock
\urldef\tempurl%
\url{https://www.youtube.com/watch?v=Axw2qu_KmWA}
\showURL{%
\tempurl}


\bibitem[Beyer et~al\mbox{.}(2020)]%
        {beyer2020we}
\bibfield{author}{\bibinfo{person}{Lucas Beyer}, \bibinfo{person}{Olivier~J
  H{\'e}naff}, \bibinfo{person}{Alexander Kolesnikov}, \bibinfo{person}{Xiaohua
  Zhai}, {and} \bibinfo{person}{A{\"a}ron van~den Oord}.}
  \bibinfo{year}{2020}\natexlab{}.
\newblock \showarticletitle{Are we done with imagenet?}
\newblock \bibinfo{journal}{\emph{arXiv preprint arXiv:2006.07159}}
  (\bibinfo{year}{2020}).
\newblock


\bibitem[cam(2019)]%
        {vid8}
\bibfield{author}{\bibinfo{person}{Highway~Traffic cam}.}
  \bibinfo{year}{2019}\natexlab{}.
\newblock
\newblock
\urldef\tempurl%
\url{https://www.youtube.com/watch?v=MNn9qKG2UFI}
\showURL{%
\tempurl}


\bibitem[Cam(2018)]%
        {vid1}
\bibfield{author}{\bibinfo{person}{Jackson Hole Wyoming USA Town Square~Live
  Cam}.} \bibinfo{year}{2018}\natexlab{}.
\newblock
\newblock
\urldef\tempurl%
\url{https://www.youtube.com/watch?v=1EiC9bvVGnk}
\showURL{%
\tempurl}


\bibitem[CAM(2021)]%
        {vid4}
\bibfield{author}{\bibinfo{person}{SOUTHAMPTON~TRAFFIC CAM}.}
  \bibinfo{year}{2021}\natexlab{}.
\newblock
\newblock
\urldef\tempurl%
\url{https://www.youtube.com/watch?v=y3NOhpkoR-w}
\showURL{%
\tempurl}


\bibitem[Canel et~al\mbox{.}(2019)]%
        {canel2019scaling}
\bibfield{author}{\bibinfo{person}{Christopher Canel}, \bibinfo{person}{Thomas
  Kim}, \bibinfo{person}{Giulio Zhou}, \bibinfo{person}{Conglong Li},
  \bibinfo{person}{Hyeontaek Lim}, \bibinfo{person}{David~G Andersen},
  \bibinfo{person}{Michael Kaminsky}, {and} \bibinfo{person}{Subramanya~R
  Dulloor}.} \bibinfo{year}{2019}\natexlab{}.
\newblock \showarticletitle{Scaling video analytics on constrained edge nodes}.
\newblock \bibinfo{journal}{\emph{arXiv preprint arXiv:1905.13536}}
  (\bibinfo{year}{2019}).
\newblock


\bibitem[Chen et~al\mbox{.}(2015)]%
        {chen2015glimpse}
\bibfield{author}{\bibinfo{person}{Tiffany Yu-Han Chen}, \bibinfo{person}{Lenin
  Ravindranath}, \bibinfo{person}{Shuo Deng}, \bibinfo{person}{Paramvir Bahl},
  {and} \bibinfo{person}{Hari Balakrishnan}.} \bibinfo{year}{2015}\natexlab{}.
\newblock \showarticletitle{Glimpse: Continuous, real-time object recognition
  on mobile devices}. In \bibinfo{booktitle}{\emph{Proceedings of the 13th ACM
  Conference on Embedded Networked Sensor Systems}}. \bibinfo{pages}{155--168}.
\newblock


\bibitem[Chin et~al\mbox{.}(2019)]%
        {chin2019adascale}
\bibfield{author}{\bibinfo{person}{Ting-Wu Chin}, \bibinfo{person}{Ruizhou
  Ding}, {and} \bibinfo{person}{Diana Marculescu}.}
  \bibinfo{year}{2019}\natexlab{}.
\newblock \showarticletitle{Adascale: Towards real-time video object detection
  using adaptive scaling}.
\newblock \bibinfo{journal}{\emph{arXiv preprint arXiv:1902.02910}}
  (\bibinfo{year}{2019}).
\newblock


\bibitem[Deshpande et~al\mbox{.}(2004)]%
        {deshpande-04}
\bibfield{author}{\bibinfo{person}{Amol Deshpande}, \bibinfo{person}{Carlos
  Guestrin}, \bibinfo{person}{Samuel~R Madden}, \bibinfo{person}{Joseph~M
  Hellerstein}, {and} \bibinfo{person}{Wei Hong}.}
  \bibinfo{year}{2004}\natexlab{}.
\newblock \showarticletitle{Model-driven data acquisition in sensor networks}.
  In \bibinfo{booktitle}{\emph{Proceedings of the Thirtieth international
  conference on Very large data bases (VLDB)}}. \bibinfo{pages}{588--599}.
\newblock


\bibitem[Du et~al\mbox{.}(2020)]%
        {du2020server}
\bibfield{author}{\bibinfo{person}{Kuntai Du}, \bibinfo{person}{Ahsan Pervaiz},
  \bibinfo{person}{Xin Yuan}, \bibinfo{person}{Aakanksha Chowdhery},
  \bibinfo{person}{Qizheng Zhang}, \bibinfo{person}{Henry Hoffmann}, {and}
  \bibinfo{person}{Junchen Jiang}.} \bibinfo{year}{2020}\natexlab{}.
\newblock \showarticletitle{Server-driven video streaming for deep learning
  inference}. In \bibinfo{booktitle}{\emph{Proceedings of the Annual conference
  of the ACM Special Interest Group on Data Communication on the applications,
  technologies, architectures, and protocols for computer communication}}.
  \bibinfo{pages}{557--570}.
\newblock


\bibitem[Emmons et~al\mbox{.}(2019)]%
        {emmons2019cracking}
\bibfield{author}{\bibinfo{person}{John Emmons}, \bibinfo{person}{Sadjad
  Fouladi}, \bibinfo{person}{Ganesh Ananthanarayanan},
  \bibinfo{person}{Shivaram Venkataraman}, \bibinfo{person}{Silvio Savarese},
  {and} \bibinfo{person}{Keith Winstein}.} \bibinfo{year}{2019}\natexlab{}.
\newblock \showarticletitle{Cracking open the dnn black-box: Video analytics
  with dnns across the camera-cloud boundary}. In
  \bibinfo{booktitle}{\emph{Proceedings of the 2019 Workshop on Hot Topics in
  Video Analytics and Intelligent Edges}}. \bibinfo{pages}{27--32}.
\newblock


\bibitem[Grois et~al\mbox{.}(2013)]%
        {grois2013performance}
\bibfield{author}{\bibinfo{person}{Dan Grois}, \bibinfo{person}{Detlev Marpe},
  \bibinfo{person}{Amit Mulayoff}, \bibinfo{person}{Benaya Itzhaky}, {and}
  \bibinfo{person}{Ofer Hadar}.} \bibinfo{year}{2013}\natexlab{}.
\newblock \showarticletitle{Performance comparison of h. 265/mpeg-hevc, vp9,
  and h. 264/mpeg-avc encoders}. In \bibinfo{booktitle}{\emph{2013 Picture
  Coding Symposium (PCS)}}. IEEE, \bibinfo{pages}{394--397}.
\newblock


\bibitem[Hancock et~al\mbox{.}(2019)]%
        {hancock2019future}
\bibfield{author}{\bibinfo{person}{Peter~A Hancock}, \bibinfo{person}{Illah
  Nourbakhsh}, {and} \bibinfo{person}{Jack Stewart}.}
  \bibinfo{year}{2019}\natexlab{}.
\newblock \showarticletitle{On the future of transportation in an era of
  automated and autonomous vehicles}.
\newblock \bibinfo{journal}{\emph{Proceedings of the National Academy of
  Sciences}} \bibinfo{volume}{116}, \bibinfo{number}{16}
  (\bibinfo{year}{2019}), \bibinfo{pages}{7684--7691}.
\newblock


\bibitem[Hinton et~al\mbox{.}(2015)]%
        {hinton2015distilling}
\bibfield{author}{\bibinfo{person}{Geoffrey Hinton}, \bibinfo{person}{Oriol
  Vinyals}, {and} \bibinfo{person}{Jeff Dean}.}
  \bibinfo{year}{2015}\natexlab{}.
\newblock \showarticletitle{Distilling the knowledge in a neural network}.
\newblock \bibinfo{journal}{\emph{arXiv preprint arXiv:1503.02531}}
  (\bibinfo{year}{2015}).
\newblock


\bibitem[Intelligence(2021)]%
        {drone_and_future}
\bibfield{author}{\bibinfo{person}{Insider Intelligence}.}
  \bibinfo{year}{2021}\natexlab{}.
\newblock \bibinfo{title}{Drone technology uses and applications for
  commercial, industrial and military drones in 2021 and the future}.
\newblock
\newblock
\urldef\tempurl%
\url{https://www.businessinsider.com/drone-technology-uses-applications}
\showURL{%
\tempurl}


\bibitem[Jingdong~Wang(2021)]%
        {visdrone}
\bibfield{author}{\bibinfo{person}{Martin~Danelljan Jingdong~Wang}.}
  \bibinfo{year}{2021}\natexlab{}.
\newblock \bibinfo{title}{VisDrone – Vision Meets Drones: A Challenge}.
\newblock
\newblock
\urldef\tempurl%
\url{{http://aiskyeye.com/}}
\showURL{%
\tempurl}


\bibitem[Kays et~al\mbox{.}(2009)]%
        {kays-09}
\bibfield{author}{\bibinfo{person}{Roland Kays}, \bibinfo{person}{Bart
  Kranstauber}, \bibinfo{person}{Patrick Jansen}, \bibinfo{person}{Chris
  Carbone}, \bibinfo{person}{Marcus Rowcliffe}, \bibinfo{person}{Tony
  Fountain}, {and} \bibinfo{person}{Sameer Tilak}.}
  \bibinfo{year}{2009}\natexlab{}.
\newblock \showarticletitle{Camera traps as sensor networks for monitoring
  animal communities}. In \bibinfo{booktitle}{\emph{2009 IEEE 34th Conference
  on Local Computer Networks}}. IEEE, \bibinfo{pages}{811--818}.
\newblock


\bibitem[Kolar et~al\mbox{.}(2016)]%
        {kolar2016ctcv}
\bibfield{author}{\bibinfo{person}{Vinay Kolar},
  \bibinfo{person}{Israat~Tanzeena Haque}, \bibinfo{person}{Vikram~P
  Munishwar}, {and} \bibinfo{person}{Nael~B Abu-Ghazaleh}.}
  \bibinfo{year}{2016}\natexlab{}.
\newblock \showarticletitle{CTCV: A protocol for coordinated transport of
  correlated video in smart camera networks}. In \bibinfo{booktitle}{\emph{2016
  IEEE 24th International Conference on Network Protocols (ICNP)}}. IEEE,
  \bibinfo{pages}{1--10}.
\newblock


\bibitem[La~Grange(2022)]%
        {vid5}
\bibfield{author}{\bibinfo{person}{Kentucky USA Virtual Railfan~LIVE
  La~Grange}.} \bibinfo{year}{2022}\natexlab{}.
\newblock
\newblock
\urldef\tempurl%
\url{https://www.youtube.com/watch?v=MDiY0SeyfGw}
\showURL{%
\tempurl}


\bibitem[Li et~al\mbox{.}(2020)]%
        {li2020reducto}
\bibfield{author}{\bibinfo{person}{Yuanqi Li}, \bibinfo{person}{Arthi
  Padmanabhan}, \bibinfo{person}{Pengzhan Zhao}, \bibinfo{person}{Yufei Wang},
  \bibinfo{person}{Guoqing~Harry Xu}, {and} \bibinfo{person}{Ravi Netravali}.}
  \bibinfo{year}{2020}\natexlab{}.
\newblock \showarticletitle{Reducto: On-camera filtering for resource-efficient
  real-time video analytics}. In \bibinfo{booktitle}{\emph{Proceedings of the
  Annual conference of the ACM Special Interest Group on Data Communication on
  the applications, technologies, architectures, and protocols for computer
  communication}}. \bibinfo{pages}{359--376}.
\newblock


\bibitem[Lin and Newley~Purnell(2019)]%
        {one_billion_cameras}
\bibfield{author}{\bibinfo{person}{Liza Lin} {and} \bibinfo{person}{The Wall
  Street~Journal Newley~Purnell}.} \bibinfo{year}{2019}\natexlab{}.
\newblock \bibinfo{title}{A World With a Billion Cameras Watching You Is Just
  Around the Corner}.
\newblock
\newblock
\urldef\tempurl%
\url{https://www.wsj.com/articles/a-billion-surveillance-cameras-forecast-to-be-watching-within-two-years-11575565402}
\showURL{%
\tempurl}


\bibitem[lot~security camera(2013)]%
        {plot}
\bibfield{author}{\bibinfo{person}{Parking lot~security camera}.}
  \bibinfo{year}{2013}\natexlab{}.
\newblock
\newblock
\urldef\tempurl%
\url{https://www.youtube.com/watch?v=U7HRKjlXK-Y}
\showURL{%
\tempurl}


\bibitem[NetEm(year)]%
        {netem}
\bibfield{author}{\bibinfo{person}{Linux NetEm}.}
  \bibinfo{year}{\the\year}\natexlab{}.
\newblock
\newblock
\urldef\tempurl%
\url{https://wiki.linuxfoundation.org/networking/netem}
\showURL{%
\tempurl}


\bibitem[of~Auburn Toomer’s Corner~Webcam(2019)]%
        {vid2}
\bibfield{author}{\bibinfo{person}{City of Auburn Toomer’s Corner~Webcam}.}
  \bibinfo{year}{2019}\natexlab{}.
\newblock
\newblock
\urldef\tempurl%
\url{https://www.youtube.com/watch?v=hMYIc5ZPJL4}
\showURL{%
\tempurl}


\bibitem[Pakha et~al\mbox{.}(2018)]%
        {pakha2018reinventing}
\bibfield{author}{\bibinfo{person}{Chrisma Pakha}, \bibinfo{person}{Aakanksha
  Chowdhery}, {and} \bibinfo{person}{Junchen Jiang}.}
  \bibinfo{year}{2018}\natexlab{}.
\newblock \showarticletitle{Reinventing video streaming for distributed vision
  analytics}. In \bibinfo{booktitle}{\emph{10th $\{$USENIX$\}$ Workshop on Hot
  Topics in Cloud Computing (HotCloud 18)}}.
\newblock


\bibitem[Pan and Yang(2009)]%
        {pan2009survey}
\bibfield{author}{\bibinfo{person}{Sinno~Jialin Pan} {and}
  \bibinfo{person}{Qiang Yang}.} \bibinfo{year}{2009}\natexlab{}.
\newblock \showarticletitle{A survey on transfer learning}.
\newblock \bibinfo{journal}{\emph{IEEE Transactions on knowledge and data
  engineering}} \bibinfo{volume}{22}, \bibinfo{number}{10}
  (\bibinfo{year}{2009}), \bibinfo{pages}{1345--1359}.
\newblock


\bibitem[Patrol(2023)]%
        {vid7}
\bibfield{author}{\bibinfo{person}{Newark Police Citizen~Virtual Patrol}.}
  \bibinfo{year}{2023}\natexlab{}.
\newblock
\newblock
\urldef\tempurl%
\url{https://cvp.newarkpublicsafety.org}
\showURL{%
\tempurl}


\bibitem[Qiu et~al\mbox{.}(2018)]%
        {qiu-18}
\bibfield{author}{\bibinfo{person}{Hang Qiu}, \bibinfo{person}{Fawad Ahmad},
  \bibinfo{person}{Fan Bai}, \bibinfo{person}{Marco Gruteser}, {and}
  \bibinfo{person}{Ramesh Govindan}.} \bibinfo{year}{2018}\natexlab{}.
\newblock \showarticletitle{AVR: Augmented Vehicular Reality}. In
  \bibinfo{booktitle}{\emph{Proceedings of the 16th Annual International
  Conference on Mobile Systems, Applications, and Services (MobiSys)}}.
  \bibinfo{pages}{81–95}.
\newblock


\bibitem[Redmon and Farhadi(2018)]%
        {incremental_improv}
\bibfield{author}{\bibinfo{person}{Joseph Redmon} {and} \bibinfo{person}{Ali
  Farhadi}.} \bibinfo{year}{2018}\natexlab{}.
\newblock \showarticletitle{YOLOv3: An Incremental Improvement}.
\newblock \bibinfo{journal}{\emph{CoRR}}  \bibinfo{volume}{abs/1804.02767}
  (\bibinfo{year}{2018}).
\newblock
\showeprint[arXiv]{1804.02767}
\urldef\tempurl%
\url{http://arxiv.org/abs/1804.02767}
\showURL{%
\tempurl}


\bibitem[Ren et~al\mbox{.}(2015)]%
        {ren2015faster}
\bibfield{author}{\bibinfo{person}{Shaoqing Ren}, \bibinfo{person}{Kaiming He},
  \bibinfo{person}{Ross Girshick}, {and} \bibinfo{person}{Jian Sun}.}
  \bibinfo{year}{2015}\natexlab{}.
\newblock \showarticletitle{Faster r-cnn: Towards real-time object detection
  with region proposal networks}.
\newblock \bibinfo{journal}{\emph{Advances in neural information processing
  systems}}  \bibinfo{volume}{28} (\bibinfo{year}{2015}).
\newblock


\bibitem[Seema and Reisslein(2011)]%
        {seema-11}
\bibfield{author}{\bibinfo{person}{Adolph Seema} {and} \bibinfo{person}{Martin
  Reisslein}.} \bibinfo{year}{2011}\natexlab{}.
\newblock \showarticletitle{Towards Efficient Wireless Video Sensor Networks: A
  Survey of Existing Node Architectures and Proposal for A Flexi-WVSNP Design}.
\newblock \bibinfo{journal}{\emph{IEEE Communications Surveys \& Tutorials}}
  \bibinfo{volume}{13}, \bibinfo{number}{3} (\bibinfo{year}{2011}),
  \bibinfo{pages}{462--486}.
\newblock
\urldef\tempurl%
\url{https://doi.org/10.1109/SURV.2011.102910.00098}
\showDOI{\tempurl}


\bibitem[Shop(2022)]%
        {vid6}
\bibfield{author}{\bibinfo{person}{Gebhardt Insurance Traffic Cam Round
  Trip~Bike Shop}.} \bibinfo{year}{2022}\natexlab{}.
\newblock
\newblock
\urldef\tempurl%
\url{https://www.youtube.com/watch?v=_XBMMTQVj68}
\showURL{%
\tempurl}


\bibitem[Simonyan and Zisserman(2014)]%
        {simonyan2014very}
\bibfield{author}{\bibinfo{person}{Karen Simonyan} {and}
  \bibinfo{person}{Andrew Zisserman}.} \bibinfo{year}{2014}\natexlab{}.
\newblock \showarticletitle{Very deep convolutional networks for large-scale
  image recognition}.
\newblock \bibinfo{journal}{\emph{arXiv preprint arXiv:1409.1556}}
  (\bibinfo{year}{2014}).
\newblock


\bibitem[Stock and Cisse(2018)]%
        {stock2018convnets}
\bibfield{author}{\bibinfo{person}{Pierre Stock} {and}
  \bibinfo{person}{Moustapha Cisse}.} \bibinfo{year}{2018}\natexlab{}.
\newblock \showarticletitle{Convnets and imagenet beyond accuracy:
  Understanding mistakes and uncovering biases}. In
  \bibinfo{booktitle}{\emph{Proceedings of the European Conference on Computer
  Vision (ECCV)}}. \bibinfo{pages}{498--512}.
\newblock


\bibitem[through Daytime(2018)]%
        {dash1}
\bibfield{author}{\bibinfo{person}{New York City~Drive through Daytime}.}
  \bibinfo{year}{2018}\natexlab{}.
\newblock
\newblock
\urldef\tempurl%
\url{https://www.youtube.com/watch?v=7HaJArMDKgI}
\showURL{%
\tempurl}


\bibitem[through Daytime(2021)]%
        {dash2}
\bibfield{author}{\bibinfo{person}{New York City~Drive through Daytime}.}
  \bibinfo{year}{2021}\natexlab{}.
\newblock
\newblock
\urldef\tempurl%
\url{https://www.youtube.com/watch?v=UPrPATfWJ2g}
\showURL{%
\tempurl}


\bibitem[Wang and Yoon(2021)]%
        {wang2021knowledge}
\bibfield{author}{\bibinfo{person}{Lin Wang} {and} \bibinfo{person}{Kuk-Jin
  Yoon}.} \bibinfo{year}{2021}\natexlab{}.
\newblock \showarticletitle{Knowledge distillation and student-teacher learning
  for visual intelligence: A review and new outlooks}.
\newblock \bibinfo{journal}{\emph{IEEE Transactions on Pattern Analysis and
  Machine Intelligence}} (\bibinfo{year}{2021}).
\newblock


\bibitem[Wang et~al\mbox{.}(2019)]%
        {wang2019bridging}
\bibfield{author}{\bibinfo{person}{Yiding Wang}, \bibinfo{person}{Weiyan Wang},
  \bibinfo{person}{Junxue Zhang}, \bibinfo{person}{Junchen Jiang}, {and}
  \bibinfo{person}{Kai Chen}.} \bibinfo{year}{2019}\natexlab{}.
\newblock \showarticletitle{Bridging the edge-cloud barrier for real-time
  advanced vision analytics}. In \bibinfo{booktitle}{\emph{11th $\{$USENIX$\}$
  Workshop on Hot Topics in Cloud Computing (HotCloud 19)}}.
\newblock


\bibitem[Zhang et~al\mbox{.}(2018)]%
        {zhang2018awstream}
\bibfield{author}{\bibinfo{person}{Ben Zhang}, \bibinfo{person}{Xin Jin},
  \bibinfo{person}{Sylvia Ratnasamy}, \bibinfo{person}{John Wawrzynek}, {and}
  \bibinfo{person}{Edward~A Lee}.} \bibinfo{year}{2018}\natexlab{}.
\newblock \showarticletitle{Awstream: Adaptive wide-area streaming analytics}.
  In \bibinfo{booktitle}{\emph{Proceedings of the 2018 Conference of the ACM
  Special Interest Group on Data Communication}}. \bibinfo{pages}{236--252}.
\newblock


\bibitem[Zhang et~al\mbox{.}(2015)]%
        {zhang2015design}
\bibfield{author}{\bibinfo{person}{Tan Zhang}, \bibinfo{person}{Aakanksha
  Chowdhery}, \bibinfo{person}{Paramvir Bahl}, \bibinfo{person}{Kyle Jamieson},
  {and} \bibinfo{person}{Suman Banerjee}.} \bibinfo{year}{2015}\natexlab{}.
\newblock \showarticletitle{The design and implementation of a wireless video
  surveillance system}. In \bibinfo{booktitle}{\emph{Proceedings of the 21st
  Annual International Conference on Mobile Computing and Networking}}.
  \bibinfo{pages}{426--438}.
\newblock


\end{thebibliography}

\appendix
\input{}

\end{document}